\documentclass[pra,twocolumn,english,showpacs,reprint,amssymb,aps, superscriptaddress, floatfix, 
longbibliography, nofootinbib]{revtex4-1}
\usepackage{graphicx,amsmath,amssymb}
\usepackage[colorlinks=true,allcolors=blue]{hyperref}
\usepackage{epstopdf}
\usepackage{bm}
\usepackage{babel}
\usepackage{ulem}
\usepackage{xcolor}

\def\sp{{\rm sp}}

\def\vj{{\bm j}}
\def\vp{{\bm p}}
\def\vd{{\bm d}}
\def\vq{{\bm q}}

\def\vA{{\bm A}}
\def\vD{{\bm D}}
\def\vr{{\bm r}}
\def\vk{{\bm k}}

\def\vq{{\bm q}}
\def\vK{{\bm K}}
\def\vQ{{\bm Q}}

\def\vJ{{\bm J}}
\def\vF{{\bm F}}
\def\vR{{\bm R}}
\def\vP{{\bm P}}

\def\cH{{\cal H}}
\def\cV{{\cal V}}
\def\cU{{\cal U}}
\def\cF{{\cal F}}
\def\cE{{\cal E}}

\def\cT{{\cal T}}

\def\cT{{\cal T}}

\begin{document}

\title{Adiabatic approach for high harmonic generation in solids induced by intense low-frequency pulses}

\author{A.~V. Flegel}

\affiliation{Department of Physics, Voronezh State University, Voronezh 394018, Russia}

\author{Liang-Wen Pi}

\affiliation{Center for Attosecond Science and Technology, Xi'an Institute of Optics and Precision Mechanics of CAS, Xi'an, Shaanxi 710119, People's Republic of China}

\author{M.~V. Frolov}

\affiliation{Department of Physics, Voronezh State University, Voronezh 394018, Russia}

\date{\today}

\begin{abstract}
An analytic description of high harmonic generation (HHG) in solids induced by intense low-frequency pulses is presented within an adiabatic approach, which treats laser-matter interactions nonperturbatively. We derive the analytical expression for the laser-dressed state of an electron in an arbitrary spatially periodic potential, taking into account multiband structure of the solid target. Closed-form formulas for electron current and HHG spectra are presented. Based on the developed theory, we provide an analytic explanation for key features of HHG yield and show that the interband mechanism of HHG prevails over the intraband one.
\end{abstract}

\maketitle

\section{Introduction}

The interaction of an intense laser pulse with matter leads to non-linear effects. The plateau effect in spectra of laser-induced processes (i.e., the weak dependence of the process probability on the number of absorbed laser photons up to the plateau cutoff) is the most intriguing and attractive. Since the first observations of plateaus in spectra of high harmonic generation (HHG) were made for rare gases~\cite{LHuillier88, HSKJPB91}, this effect has been extensively studied both experimentally and theoretically for atomic targets and served as a base for the production of extremely short attosecond pulses that led to attosecond science~\cite{SIKVJPB06, KIRMP09}. A clear physical explanation of the plateau structures can be given within the rescattering picture~\cite{SYDKPRL93, CorkumPRL93, PBNWJPB94}. Under the action of intense laser field, an atomic electron tunnels into a laser-modified continuum, where it propagates along a closed classical trajectory until the moment of rescattering (i.e., the return to the parent ion). At the rescattering moment, the electron recombines with the parent ion, emitting a high-order harmonic photon.

The efficiency of HHG for a gaseous medium is suppressed by low density, while for a condensed medium it is significantly enhanced by a larger number of active charges that affect the laser-induced current and the response of the medium. For the first time, HHG in a bulk crystal was observed in Ref.~\cite{Ghimire2011}, which led to a dramatic increase in interest in HHG for different condensed materials (see reviews~\cite{KKYRMP18, Park22, Goulielmakis2022, Yue2022, Ghimire2019, LiRPP2023} and references therein). Despite the great (resulted in a large number of papers) interest in HHG by solids and HHG-based creation of ultrashort pulses, a complete understanding of the physics of this phenomenon has not yet been achieved because of the complex structure of solids, which leads to complex laser-induced dynamics.  The generally accepted mechanisms of HHG in solids consist of intraband current~\cite{Ghimire2011,GhimirePRA12} and interband polarization~\cite{VampaPRL14,Vampa2015}. Both mechanisms are based on the tunneling of an electron into the conduction band, followed by the dynamics of the formed electron--hole pair in an intense laser field. The intraband mechanism involves the movement of a hole (in the valence band) and an electron (in the conduction band) with an anharmonic temporal dependence due to the nonparabolic dependence of energy on the laser-modified crystal momentum. The interband mechanism involves electron-hole recombination with the emission of a harmonic photon, similar to the rescattering scenario for HHG by a single atom. 

Although the above-mentioned quasiclassical HHG-mechanisms agree with both experimental observations and the results of other theoretical approaches, the developing of a consistent analytical approach to describe HHG in solids is still relevant. Such an approach assumes an accurate nonperturbative account of both interactions of an active electron with the laser pulse and with the periodic potential of a crystal lattice. The most exact accounting of these interactions and the effects of the material characteristics can be achieved using numerical methods such as based on the time-dependent Sch\"{o}dinger equation (TDSE), semiconductor Bloch equations (SBEs), and time-dependent density functional theory (see, e.g., Refs.~\cite{LiPRL19,GoldePRB08,LuuPRB16,TancognePRL17,Hansen17,Yu19}). Although numerical approaches provide ab initio results, their ability is quite restricted by the intensity, wavelength, and wave form of a laser pulse and presents a time-consuming task for an intense tailored laser pulse having carrier frequencies in the midinfrared range. We note that numerical results cannot provide a fundamental insight into a physical problem, since they are obtained for particular values of parameters. Moreover, some applications in physics require a robust analytical expression for probability of a fundamental process in an intense laser field with accuracy comparable with time-consuming numerical ab initio results. The solid analytical expressions can be obtained within analytical methods, whose implementation in physical problems is extremely important. 

For laser-induced HHG by atoms, the mostly accepted approach is the strong-field approximation (SFA) and consists in the Born-like expansion of the HHG-amplitude in the formal series in an atomic potential~\cite{MEAdv03}. The expansion terms can be analyzed within the saddle-point method, resulting in the quantum orbit approach~\cite{SalScience01}, which provides a transparent physical interpretation of strong-field phenomena in terms of classical trajectories and constitutes a background for a parametrization of the HHG-amplitude in terms of a product of laser and atomic parameters~(see, e.g., Ref.~\cite{LZMCVPRL07}). An alternative to SFA-based approaches is adiabatic theory~\cite{OTMPRA12,FlegPRA21,FrolovJPA2023}, which operates with the smallness of the laser-pulse carrier frequency with respect to both the electron quiver energy in a strong low-frequency field and the atomic ionization potential. In contrast to the SFA, the adiabatic approximation provides an accurate account of an atomic potential, whose influence on strong-field processes may be crucial~\cite{FMSERSPRL09, FMSPRA10, NatureTralerro, SSetalJPB12, SilaevPRA20, Romanov2021}.

In this work, we develop an analytic approach to study the HHG process in solids based on the key provisions of the adiabatic theory of the laser-induced atomic processes. In contrast to atomic processes, we use the accelerated Bloch states~\cite{Krieger86} for an electron in the periodic potential as a basis for the laser-dressed state instead of the Volkov states for a free electron. We show that the wave function of an electron in the periodic potential and intense low-frequency laser field can be presented as a sum of the adiabatic part, which exactly takes into account the instantaneous couplings of all crystal bands, and the part responsible for the interband transitions. The analytical structure of the wave function is utilized to obtain the factorization for the laser-induced current and HHG-amplitude in terms of laser and matter parameters in a similar manner as for HHG in gaseous media. Using our analytical results, we discuss the physical picture and mechanisms of HHG in solids.

\section{Theoretical approach}
\subsection{General equations}
\label{GenEq}

In this section, we briefly present general equations describing an electron current in a crystal subjected to homogeneous intense IR field within the one active electron approximation (some additional details can be found in Ref.~\cite{LiRPP2023}). The wave function $\Psi_{\vq}(\vr,t)$ of an active electron with crystal momentum $\vq$ (i.e. quasi-momentum) in the IR field with vector potential $\vA(t)$ satisfies the time-dependent Schr\"{o}dinger equation (TDSE):
\begin{eqnarray}
\label{TDSE}
&& i\frac{\partial \Psi_\vq(\vr,t)}{\partial t}= \hat{H}\Psi_\vq(\vr,t),\\
\label{Hamiltonian:IR}
&& \hat{H}=\frac{1}{2}[\hat{\vp}+\vA(t)]^2+U(\vr),
\end{eqnarray}
where $\hat{\vp}= -i\nabla$ is the momentum operator, $U(\vr)=U(\vr+\vR)$ is the periodic crystal potential (for all $\vR$ from the Bravais lattice).

We seek the solution of Eq.~\eqref{TDSE} in the form of expansion over the set of accelerated Bloch states or Houston functions $\varphi_{\vq,n}(\vr,t)$~\cite{Krieger86}:
\begin{equation}
\label{Psi-Bloch}
\Psi_\vq(\vr,t) = \sum_{n}a_{\vq,n}(t)\varphi_{\vq,n}(\vr,t).
\end{equation}
The wavefunctions $\varphi_{\vq,n}(\vr,t)$ are eigenstates of the time-dependent (instantaneous) Hamiltonian $\hat{H}$,
\begin{eqnarray}
\label{Houston-Eq}
&& \hat{H}\varphi_{\vq,n}(\vr,t) = E_n(t)\varphi_{\vq,n}(\vr,t),\\ 
\nonumber
&& E_n(t)\equiv E_n[\vQ(t)],
\end{eqnarray}
and can be expressed in terms of field-free states $\psi_{\vq,n}(\vr)$ with energies $E_n(\vq)$ of the $n$th band taken at the instantaneous field-dressed quasi-momentum $\vQ(t)= \vq+\vA(t)$ [i.e., replacing $\vq\to\vQ(t)$]:
\begin{equation}
\label{Houston-psi}
\varphi_{\vq,n}(\vr,t) = e^{-i\vr\vA(t)}\psi_{\vQ(t),n}(\vr) = e^{i\vq\vr}u_{\vQ(t),n}(\vr),
\end{equation}
where $u_{\vQ(t),n}(\vr) = u_{\vQ(t),n}(\vr+\vR)$.

The time-dependent coefficients $a_{\vq,n}(t)$ in Eq.~\eqref{Psi-Bloch} satisfy the system of differential equations:
\begin{equation}
\label{a(t)-Eq}
i\dot{a}_{\vq,n}(t) = E_n(t){a}_{\vq,n}(t) + \sum_{n'}\cV_{nn'}(t){a}_{\vq,n'}(t),
\end{equation}
where $\cV_{nn'}(t)=-\langle\varphi_{\vq,n}|i\partial_t|\varphi_{\vq,n'}\rangle$ is a nonadiabatic coupling of adiabatic states $\varphi_{\vq,n}(\vr,t)$ and $\varphi_{\vq,n'}(\vr,t)$:
\begin{eqnarray}
\label{Xnn'}
&& \cV_{nn'}(t) = \langle u_{\vQ(t),n}|\hat\cV|u_{\vQ(t),n'}\rangle = \vF(t)\vd_{nn'}[\vQ(t)], \\
\nonumber 
&&\hat\cV \equiv\hat\cV(t) = i\vF(t)\cdot\nabla_\vq,\\ 
\label{dnn'}
&& \vd_{nn'}[\vQ(t)] = \langle u_{\vQ(t),n}|i\nabla_\vq| u_{\vQ(t),n'} \rangle,
\end{eqnarray}
and $\vF(t)=-\dot{\vA}(t)$ is the electric vector of the laser field.  We assume that the electron is initially in the valence band state $\psi_{\vq,0}$, so that Eqs.~\eqref{a(t)-Eq} are accompanied by the initial condition at the time $t_0$ (in general $t_0=-\infty$) of turn-on of the laser field:
\begin{equation}
\label{a(t)-ic}
a_{\vq,n}(-\infty)=\delta_{n0}.
\end{equation}
It can be shown that system of equations~\eqref{a(t)-Eq} leads to the semiconductor Bloch equations (SBEs) by introducing the density matrix $\rho_{\vq,nn'}(t)={a}_{\vq,n}(t){a}^*_{\vq,n'}(t)$~\cite{LiRPP2023}:
\begin{eqnarray}
\label{SBE}
\nonumber
&& i\dot{\rho}_{\vq,nn'}(t) = [E_n(t)-E_{n'}(t)]\rho_{\vq,nn'}(t) \\
&& \quad + \sum_{m}[\cV_{nm}(t)\rho_{\vq,mn'}(t)-\rho_{\vq,nm}(t)\cV_{mn'}(t)].
\end{eqnarray}

The HHG amplitude for the harmonic frequency $\Omega$ is determined by the Fourier component $\vJ(\Omega)$ of the current density $\vj(t)$:
\begin{equation}
\label{J(Omega)}
\vJ(\Omega) = \int e^{i\Omega t}\vj(t)dt,\quad \vj(t) = \int \vj_\vq(t) \frac{d\vq}{(2\pi)^3},
\end{equation}
where $\vj_\vq(t)$ is the contribution of the current from the state with given quasi-momentum~$\vq$:
\begin{equation}
\label{j_q}
\vj_\vq(t)= -\langle \Psi_\vq(t)|[\hat\vp+\vA(t)]| \Psi_\vq(t)\rangle,
\end{equation} 
and $\vj(t)$ is the total current density averaged over the $\vq$-space. 
Using the expansion~\eqref{Psi-Bloch} for the state $\Psi_\vq$, from Eqs.~\eqref{j_q} we obtain the following two-term expression for~$\vj(t)$:
\begin{eqnarray}
\label{j-sum}
&&\vj(t) = \vj_a(t)+\vj_e(t),\\
\label{ja}
&&\vj_a(t) = - \sum_n\int \frac{d\vq}{(2\pi)^3} \, \rho_{\vq,nn}(t) \nabla_\vq E_n[\vQ(t)],\\
\label{je}
\nonumber
&&\vj_e(t) = - i{\sum_{n,n'}}'\int \frac{d\vq}{(2\pi)^3}\, \rho_{\vq,nn'}(t) \\ 
&&\phantom{vj_e(t)}\times \left[
E_{n'}(t)-E_n(t)\right]\vd_{n'n}[\vQ(t)],
\end{eqnarray}
where the symbol ${\sum}'$ means the summation over $n\neq n'$. The term $\vj_a$ is expressed through the populations $\rho_{nn}=|a_{\vq_n}|^2$ of all particular bands and describes the intraband current, while the term $\vj_e$ contains the interband polarization $\vd_{n'n}$ and gives the interband current.

\subsection{Adiabatic analysis of the electron laser-dressed state with a given quasi-momentum}
\label{Sec:adiabat-IR}

We develop our analytic approach for the following assumptions. We consider a low-frequency laser field interacting with a semiconductor or dielectric with band energies $E_n$ and band gaps $\Delta_{nn'}=E_n-E_{n'}$ much larger than laser-pulse carrier frequency and interband couplings $\cV_{nn'}$:
\begin{subequations}
\label{conditions}
\begin{eqnarray}
\label{cond1}
&&|E_{n}-E_{n'}|\gg\omega,\\
\label{cond2}
&&|E_{n}-E_{n'}|\gg|\cV_{nn'}|,\quad |E_n|\gg |\cV_{nn}|.
\end{eqnarray}
\end{subequations}
The first condition~\eqref{cond1} means that laser-induced electron transitions between different bands are described by the tunneling mechanism in the instantaneous field with exponentially small Zener-type transition amplitudes~\cite{Zener34}. The second inequality~\eqref{cond2} is less strict for the adiabatic consideration, but allows one to use perturbative analysis with respect to the matrix elements $\cV_{nn'}\sim FR$ (where $F$ and $R$ are the characteristic field strength and the lattice period, respectively). The condition~\eqref{cond2} also means that the Berry phase $\phi_n^{(\rm B)}(t)=-\int^t \cV_{nn}(\tau) d\tau$ gives the factor $\exp[i\phi^{(\rm B)}(t)]$, which should be treated as a slow varying function in time, in contrast to the rapidly oscillating dynamic factor $\exp[-i\int^t E_n(\tau) d\tau]$.

Within inequality~\eqref{cond2}, we rewrite Eq.~\eqref{a(t)-Eq} in the following iterative form:
\begin{subequations}
\label{a-PT}
\begin{eqnarray}
&&a_{\vq,n}(t) = \sum_{m=0}^{\infty}a^{(m)}_{\vq,n}(t),\\
&&a^{(0)}_{\vq,n}(t) = \delta_{n0}e^{-i\int^t E_0(\tau)d\tau},\\
\nonumber
&&a^{(m)}_{\vq,n}(t) = -i\sum_{n'}\int_{-\infty}^t dt' e^{-i\int^t_{t'} E_n(\tau)d\tau} \\
&&\phantom{a^{(m)}_{\vq,n}(t) =} \times \cV_{nn'}(t')a^{(m-1)}_{\vq,n'}(t').
\label{a-rec}
\end{eqnarray}
\end{subequations}

In the low-frequency limit (which is appropriate for the mid-infrared region), the preexponential factor $\cV_{nn'}(t')$ can be considered as a slowly varying function of $t'$. Thus, in accordance with the key provisions of the adiabatic approach, the main contribution to the integral in Eq.~\eqref{a-rec} is given by the well-separated vicinities of the point $t'=t$ and the set of saddle points $t'=t'_s{}^{(\sp)}$ of the exponential phase in the integrand. The detailed adiabatic analysis of Eqs.~\eqref{a-PT} is given in Appendix~\ref{app:1} and we proceed our analysis with the adiabatic result for $a_{\vq,n}(t)$. 

The solution of Eqs.~\eqref{a-PT} can be written as a sum of two terms:
\begin{equation}
\label{aqn:0+sp}
a_{\vq,n}(t) = a^0_{\vq,n}(t) +a^\sp_{\vq,n}(t),
\end{equation}
where the first term $a^0_{\vq,n}(t)$ takes into account instantaneous couplings $\cV_{nn'}(t)$ between bands and originates from the contribution of the upper limit $t'=t$ for temporal integrals in Eq.~\eqref{a-rec} [see Eq.~\eqref{0:a:fin} in Appendix~\ref{app:1}], the second term $a^\sp_{\vq,n}(t)$ originates from the contribution of saddle points $t'=t'_s{}^{(\sp)}$ and involves tunneling transitions from the valence band to different conduct bands [see Eq.~\eqref{sp:a:fin} in Appendix~\ref{app:1}]. Substituting the expressions obtained for $a_{\vq,n}(t)$ into Eq.~\eqref{Psi-Bloch}, we express the laser-dressed state $\Psi_\vq$ in the form: 
\begin{equation}
\label{IR-dressed:final}
\Psi_\vq(\vr,t) = \sum_n c_n e^{-i\int^t \cE_n(\tau)d\tau}\Phi_{\vq,n}(\vr,t),
\end{equation}
where $c_0=1$, coefficients $c_n$ for $n>0$ represent the sum of partial $s$-terms associated with different tunneling events:
\begin{subequations}
\label{coeffs}
\begin{eqnarray}
\label{c:nu}
\nonumber
&&c_n = \sqrt{\frac{2\pi}{i}}\sum_s e^{i\int^{t'_s}\left[E_n(\tau)-\cE_0(\tau)\right]d\tau}  \\
&&\quad \times \frac{e^{-\frac{\varkappa^3_{n 0}}{3\cF_{n 0}} }}{\sqrt{\cF_{n 0}\varkappa_{n 0}}} \langle u_{\vQ(t'_s),n}|\hat\cV(t'_s)|\cU_{\vQ(t'_s),0}\rangle,\\
\label{varkappa}
&&\varkappa_{n0} = \sqrt{2\Delta_{n0}(t_s')},\\ 
\label{calF}
&&\cF_{n0} = \sqrt{\vF^2(t'_s)\nabla_\vq^2\Delta_{n0}(t_s') - \dot{\vF}(t'_s)\nabla_\vq\Delta_{n0}(t_s')},\;\;\;\;\;
\end{eqnarray}
\end{subequations}
and \textit{real} tunneling times $t_s'$ are solutions of the saddle-point equation: 
\begin{equation}
\label{speq-real}
{\vF}(t'_s)\cdot\nabla_\vq\Delta_{n0}(t_s') = 0,
\end{equation}
where $\Delta_{n0}(t) = E_n(t)-E_0(t)$ is the gap between the valence and conduction bands. 

In Eq.~\eqref{IR-dressed:final}, the wave functions $\Phi_{\vq,n}(\vr,t)$ are electron states for different bands [i.e., for valence ($n=0$) and conduct ($n\neq 0$) bands] having energies $\cE_n(t)\equiv \cE_n[\vQ(t)]$ with exact account of instantaneous interband coupling:
\begin{eqnarray}
\label{Phi-U}
&& \Phi_{\vq,n}(\vr,t)=e^{i\vq\vr}\cU_{\vQ(t),n}(\vr),\\
\label{cE}
&& \cE_n(t) = E_n(t) + \langle u_{\vQ(\tau),n}|\hat{\cV}(\tau)|\cU_{\vQ(\tau),n}\rangle,
\end{eqnarray}
where periodic functions $\cU_{\vQ(t),n}(\vr)$ [$\cU_{\vQ(t),n}(\vr+\vR)=\cU_{\vQ(t),n}(\vr)$] satisfy the equation [cf. Eq.~\eqref{app:cU}]
\begin{eqnarray}
\label{cU}
\nonumber
&& |\cU_{\vQ(t),n}\rangle = \sum_{m=0}^{\infty}\big[G'_{E_n(t)}\hat\cV(t)\big]^m |u_{\vQ(t),n}\rangle\\
&& \phantom{|\cU_{\vQ(t),n}\rangle} =|u_{\vQ(t),n}\rangle + G'_{E_n(t)}\hat\cV(t)|\cU_{\vQ(t),n}\rangle.    
\end{eqnarray}
Here we introduce the reduced ``stationary'' Green's function $G'_{E_n(t)}\equiv G'_{E_n(t)}(\vr,\vr';t)$ for an electron in the periodic potential $U(\vr)$, which involves $t$ as a parameter:
\begin{equation}
\label{Green-red}
G_{E_n(t)}' = {\sum_{\nu\neq n}}\frac{|u_{\vQ(t),\nu}\rangle\langle u_{\vQ(t),\nu}|}{E_{n}(t) - E_\nu(t)}.
\end{equation}

In Eq.~\eqref{IR-dressed:final}, the zeroth term $\Psi_{\vq}^0$, 
\begin{equation}
\label{Psi00}
\Psi_{\vq}^0=e^{-i\int^t \cE_0(\tau)d\tau}\Phi_{\vq,0},
\end{equation}
originates from the term $a^0_{\vq,n}(t)$ in Eq.~\eqref{aqn:0+sp} and describes the electron state in the original valence band resulting of the laser-induced instantaneous interaction with all other conduction bands. This state is similar to the atomic adiabatic laser-dressed state, which can be represented by the part of the atomic wave function (analytical with respect to the laser strength) in the instantaneous laser field~\cite{FrolovJPA2023}. In fact, the wave functions $\Phi_{\vq,n}(\vr,t)$ are presented by series expansions in the field strength $\vF(t)$ [through the matrix elements $\cV_{nn'}$, cf.~\eqref{Xnn'}], which are assumed to be convergent and thus $\Phi_{\vq,n}(\vr,t)$ are also analytical functions with respect to $\vF(t)$. Moreover, the time dependence of the zeroth term $\Psi^0_{\vq}$ is determined mainly by the exponential $\exp[-i\int^t \cE_0(\tau)d\tau]$ and does not contain electron tunneling transitions to excited conduction bands. For this reason, it represents a ``slow'' part of the laser-dressed state (as a function of time). 

All other terms with $n>0$ in Eq.~\eqref{IR-dressed:final} form the electron wave packet $\Psi_{\vq}^\sp$,
\begin{equation}
\label{Psi:sp}
\Psi_{\vq}^\sp=\sum_{n>0} c_n e^{-i\int^t \cE_n(\tau)d\tau}\Phi_{\vq,n},
\end{equation}
containing the adiabatic states $\Phi_{\vq,n}$ in the excited conduction bands. According to the explicit analytical expression for the coefficients $c_n$, given by Eq.~\eqref{c:nu}, the states $\Phi_{\vq,n}$ are formed by two steps: (i) tunneling of the electron from the initial valence band to the conduction $n$th band at moment of time $t'_s$ and (ii) propagation of the electron up to time $t$ with instantaneous quasi-momentum change from $\vQ(t_s')$ to $\vQ(t)=\vQ(t_s') - \vA(t'_s) + \vA(t)$. The wave packet $\Psi_{\vq}^\sp$ can be considered as a fast part of the laser-dressed state $\Psi_\vq$ for the crystal electron similarly to the case of an atomic electron.\footnote{The terms ``slow'' and ``fast'' states are conditional and should be understood in the sense of an adiabatic approximation for the electron state in a low-frequency laser field.} For the latter case, the role of the excited states $\Phi_{\vq,n}$ is played by electron states in the laser-dressed continuum. 

Tunneling times $t'_s$ are solutions of Eq.~\eqref{speq-real}, and describe the transition time moments for an electron from the valence band to the conduction band. For $t_s'$, we do not take into account the contribution of instantaneous coupling $\hat{\cV}(t)$ (which in the lowest perturbation order is determined by the Berry phase $\phi_n^{(B)}$) due to the relative smallness of its derivative. The second simplification of our approach is an account of the tunneling factors in the lowest non-vanishing order. To correct those results, one should evaluate the integrals in Eq.~\eqref{app:a-recur} over two different time variables using the saddle-point method and take into account the corresponding saddle-point contributions. Such an approach leads to the cascading mechanism of electron transition to the excited bands by double tunneling and population of an intermediate band. It is not considered in this paper and will be investigated elsewhere. 

In conclusion of this section, we note that within the adiabatic approximation, the states $\Phi_{\vq,n}(\vr,t)$ with energies $\cE_n(t)$ form a set of eigenfunctions and eigenvalues of the instantaneous Hamiltonian $\hat{\cH}(t)=\hat{H}-i\partial_t$: 
\begin{eqnarray}
\label{eigenproblem:cV}
&&\hat{\cH}(t)\Phi_{\vq,n}(\vr,t) = \cE_n(t)\Phi_{\vq,n}(\vr,t),\\
\nonumber
&&\hat{\cH}(t) = \frac{1}{2}[\hat{\vp}+\vA(t)]^2+U(\vr) - i\frac{\partial}{\partial t}.
\end{eqnarray}
This result can be used to simplify some derivations as well as to build adiabatic Green's function on the basis of states $\Phi_\vq$ for the electron in the crystal potential subjected to an intense low-frequency laser field.

\subsection{Analytical expressions of the HHG-amplitude}

With a known analytical structure of the laser-dressed electron state, we derive analytical expressions for the laser-induced current according to Eqs.~\eqref{J(Omega)} and~\eqref{j_q}. Following the results of Sec.~\ref{Sec:adiabat-IR}, we write the current $\vj_\vq(t)$ in the laser-dressed state $\Psi_\vq$ with a given momentum $\vq$ as a sum of four matrix elements:
\begin{eqnarray}
\label{j-4}
\nonumber
&&\vj_\vq(t) = -\langle \Psi_\vq^0(t)|\hat\vP(t)|\Psi_\vq^0(t)\rangle - \langle \Psi_\vq^0(t)|\hat\vP(t)|\Psi_\vq^\sp(t)\rangle\\
&&\quad - \langle \Psi_\vq^\sp(t)|\hat\vP(t)|\Psi_\vq^0(t)\rangle - \langle \Psi_\vq^\sp(t)|\hat\vP(t)|\Psi_\vq^\sp(t)\rangle.
\end{eqnarray}
With the result~\eqref{IR-dressed:final}, the matrix elements in Eq.~\eqref{j-4} can be expressed in terms of the matrix elements $\vP_{nn'}$ based on the adiabatic functions $\Phi_{\vq,n}$:
\begin{eqnarray}
\label{Pnn}
\nonumber
&& \vP_{nn'}(t) = \langle\Phi_{\vq,n}(t)|\hat{\vP}(t)|\Phi_{\vq,n'}(t)\rangle \\
\nonumber
&& \phantom{\vP_{nn'}(t)} = -i\langle\Phi_{\vq,n}(t)|[\vr\hat{\cH}-\hat{\cH}\vr]|\Phi_{\vq,n'}(t)\rangle=\\
&& \phantom{\vP_{nn'}(t)} = i[\cE_n(t)-\cE_{n'}(t)]\vD_{nn'}(t)+\nabla_\vq\cE_n(t)\delta_{nn'},\;\;\;\;\;\;\;\;
\end{eqnarray}
where
\begin{equation}
\label{Dnn}
\vD_{nn'}(t)\equiv \vD_{nn'}[\vQ(t)]=\langle \cU_{\vQ(t),n}|i\nabla_\vq| \cU_{\vQ(t),n'} \rangle.\;\;\;
\end{equation}
We note that for moderate laser intensities in the IR region one can neglect the effects of the adiabatic coupling $\hat\cV(t)$ in the states $\Phi_{\vq,n}$, so that $\Phi_{\vq,n}\approx\varphi_{\vq,n}$ and $\vD_{nn'}\approx\vd_{nn'}$ [cf. Eq.~\eqref{dnn'}].

Using Eqs.~\eqref{Psi00} and~\eqref{Pnn} for the first term in Eq.~\eqref{j-4}, we obtain
\begin{equation*}
\label{j1}
\langle \Psi_\vq^0(t)|\hat\vP(t)|\Psi_\vq^0(t)\rangle = -\nabla_\vq\cE_0(t).
\end{equation*}
In particular, for an isotropic dependence $\cE_0[\vQ(t)]$ on the momentum $\vQ(t)$, this term tends to zero after integrating over the $\vq$-space. The second and third terms in Eq.~\eqref{j-4} are time-reversed to each other. The second term describes the tunneling of electron from an initial band to an excited band as the first step of the scenario for induction of the current $\vj_\vq(t)$, while for the third, the tunneling occurs after propagation. The inverse scenario (described by the third term) is similar to the HHG process in atomic gases and sufficiently suppressed.  Finally, the fourth term in Eq.~\eqref{j-4} involves two tunneling factors and should be omitted within the approximation used. 

For these reasons, only the second matrix element in Eq.~\eqref{j-4} will be considered further. As a result, we have:
\begin{eqnarray}
\label{j-fin}
\nonumber
&& \vj_\vq(t) \approx \vj_{\vq,e} = i\sum_{n\neq 0} c_n e^{i\int^t[\cE_0(\tau)-\cE_n(\tau)]d\tau}\\
&& \phantom{\vj_\vq(t) \approx \vj_{\vq,e}} \times[\cE_n(t)-\cE_0(t)]\vD_{0n}(t),
\end{eqnarray}
where coefficients $c_n$ are given in Eq.~\eqref{c:nu}. For laser fields of moderate intensities, we can neglect the effects of instantaneous (adiabatic) couplings in the laser-induced current. In this case, the energy $\cE_n$ and polarization matrix $\vD_{0n}$ can be respectively replaced by $E_n$ and $\vd_{0n}$ in Eq.~\eqref{j-fin}. Comparing this approximated result with Eq.~\eqref{je}, we conclude that the main mechanism for current induction is determined by the interband polarization $\vd_{0n}$ between the zeroth and $n$th bands. 

The intraband current $\vj_{a}$ in Eq.~\eqref{ja} is negligibly small within the approximation used. Indeed, as follows from Eq.~\eqref{j-4}, the intraband current $\vj_{\vq,a}$ for a given $\vq$ originates from (i) the first term, which describes the current in the zeroth band (i.e., a current of a hole in the valence band), (ii) the fourth term, which includes the sum the partial currents in different conduct bands with $n>0$ multiplied by the product of two exponentially small tunneling factors from the zeroth band to the $n$th band:
\begin{equation}
\label{j-fin:a}
\vj_{\vq,a}(t) = -\sum_{n}|c_n|^2\nabla_\vq\cE_n(t).
\end{equation}

To find analytical results for the HHG-yield, we substitute the result~\eqref{j-fin} in Eqs.~\eqref{J(Omega)} and estimate the integrals over $t$ and $\vq$ by the saddle-point method. The saddle-point equations are:
\begin{subequations}
\label{HHG:sp-Eq}
\begin{eqnarray}
\label{sp-Eq:t}
&& \Delta_{n0}[\vQ(t_s)] = E_n(t_s) - E_0(t_s) = \Omega,\\
\label{sp-Eq:q}
&& \nabla_\vq\int_{t'_s}^{t_s}\Delta_{n0}[\vQ(\tau)]d\tau = 0.
\end{eqnarray}
\end{subequations}
The Eq.~\eqref{sp-Eq:t} describes the electron-hole annihilation (recombination) accompanied by harmonic emission at time $t=t_s$, while Eq.~\eqref{sp-Eq:q} determines the most contributing crystal momentum $\vq=\vk_s\equiv\vk(t_s,t_s')$. The dependence of this crystal momentum $\vk_s$ on the tunneling ($t_s'$) and recombination ($t_s$) times means that Eqs.~\eqref{sp-Eq:t} and~\eqref{speq-real} compose the coupled equation system. For convenience, we mark the couples of corresponding times $(t_s,t_s')$ by the same index $s$.
As a result, the current $\vJ(\Omega)$ at harmonic frequency $\Omega$ can be presented in the following form:
\begin{equation}
\label{J(Omega)-fin}
\vJ(\Omega) = \Omega\sum_{s}\sum_{n\neq 0} c^{(\rm tun)}_{n0,s} c^{(\rm pr)}_{n0,s}\vD_{0n}(\vK_s),
\end{equation}
where $\vK_s = \vk_s+\vA(t_s)$. The dipole matrix element $\vD_{0n}(\vK_s)$ in Eq.~\eqref{J(Omega)-fin} describes the recombination step of the 3-step scenario for the interband HHG in a crystal. Tunneling factors ($c^{(\rm tun)}_{n0,s}$) and propagation factors ($c^{(\rm pr)}_{n0,s}$) are defined as:
\begin{eqnarray}
&& c^{(\rm tun)}_{n0,s} = \frac{e^{-\frac{\varkappa^3_{\nu 0}}{3\cF_{\nu 0}} }}{\sqrt{\cF_{\nu 0}\varkappa_{\nu 0}}} \langle u_{\vq,n}|\hat\cV(t'_s)|\cU_{\vq,0}\rangle\Bigg|_{\vq= \vK'_s},\\
&& c^{(\rm pr)}_{n0,s} = - \frac{e^{iS(t_s,t'_s)}}{\sqrt{2\pi i \cT^3_s\alpha_s} },
\end{eqnarray}
where $\vK'_s = \vk_s+\vA(t'_s)$,
\begin{eqnarray*}
&& S(t_s,t'_s)=\int^{t_s}_{t'_s}\left[\cE_{0}(\tau)-E_n(\tau)\right]d\tau-\int^{t_s}\Delta E_{n}(\tau)d\tau,\\
&& \alpha_s = \left[\vF(t_s)+\frac{1}{\cT_s}\nabla_\vq \Delta_{n0}(\vq)\right]\cdot\nabla_\vq \Delta_{n0}(\vq)\Bigg|_{\vq= \vK_s},\\
&& \cT_s=\nabla_\vq^2 \int_{t_s'}^{t_s}\Delta_{n0}[\vq +\vA(\tau)] d\tau\Bigg|_{\vq= \vk_s},\\
\end{eqnarray*}
and $\Delta E_{n}(\tau)$ is the energy shift of $n$th band caused by instantaneous couplings between all bands [cf. Eq.~\eqref{app:cE}].




\section{Summary}

In this work, we provide the theoretical approach to study the analytical structure of the wave function $\Psi_{\vq}(\vr,t)$ for an electron with a given crystal momentum $\vq$ interacting with a spatially periodic potential $U(\vr)$ and an intense low-frequency laser pulse. Our adiabatic approach treats the electron-laser interaction quasi-classically and exactly takes into account effects of the electron-matter interaction (through the potential $U(\vr)$). The closed-form analytical expression for $\Psi_{\vq}(\vr,t)$ represents a superposition of laser-dressed states for an electron in different $n$-th bands $\Phi_{\vq,n}(\vr,t)$ [see Eq.~\eqref{IR-dressed:final}], which in turn describe the electron state in the potential $U(\vr)$ and the static electric field with the strength equal to the instantaneous value $\vF(t)$ of the laser field. It should be emphasized that the developed adiabatic theory allows one to take into account an arbitrary number of instantaneous interband couplings, so that the state $\Phi_{\vq,n}(\vr,t)$ describes the effect of the band structure exactly. The weight (or amplitude) of the state $\Phi_{\vq,n}(\vr,t)$ in the superposition for the resulting laser-dressed state $\Psi_{\vq}(\vr,t)$ is determined by the sum over all different tunneling events (at times $t'_s$) from an initial valence band to $n$-th conduct band and is proportional to the Zener-type tunneling exponential [cf. Eq.~\eqref{c:nu}]. We note the following remarkable analogy between the results for the analytical structure of the state $\Psi_{\vq}(\vr,t)$ and the laser-dressed state of an atomic electron~\cite{FlegPRA21,FrolovJPA2023}. In both cases the wave function can be presented as a sum of two terms: ``slow'' and ``fast'', where ``slow'' part of the atomic state is given by the atomic electron wave function in the instantaneous field $\vF(t)$, and the ``fast'' part represents the superposition of the scattering states in the laser-modified continuum. For the crystal electron, the ``slow'' part is given by the wave function $\Phi_{\vq,0}(\vr,t)$ in the initial valence (0-th) band, while the ``fast'' part involves the superposition of all laser-modified states of conduct bands (instead of continuum states for the atomic case). 

Based on the analytical expressions for the wave functions, we have obtained the parameterization for the electron current density $\vJ(\Omega)$ in the crystal at a given frequency $\Omega$, the square of which determines the HHG spectra. Within the adiabatic conditions, assuming a large band gap with respect to the carrier frequency of the laser pulse, we found that the main mechanism for HHG is interband polarization, while the intraband current sufficiently depends on the symmetry properties of the dispersion relation $E_n(\vq)$. For the interband mechanism, the analytic result for $\vJ(\Omega)$ is given by Eq.~\eqref{J(Omega)-fin}, representing the current $\vJ(\Omega)$ as a sum of partial terms, each of them has clear physical ``three-step'' parameterization in terms of laser and crystal parameters. These partial HHG amplitudes originate from different tunneling events from the valence band to one of the conduct bands, involve the electron-hole propagation factor as a second step of the HHG mechanism, and the polarization dipole factor describing the third step --- radiative annihilation. Since tunneling to the lowest conduct band is more probable than to the higher excited bands in the adiabatic limit, the two-band model is a reasonable model for the process considered. However, the presented approach allows one to take into account an arbitrary number of the higher bands as well as to analyze the contribution of the instantaneous interband couplings in arbitrary high order.  

\acknowledgments

AVF thanks for hospitality of Xi'an Institute of Optics and Precision Mechanics (XIOPM) of Chinese Academy of Sciences (the CAS PIFI programm), where part of this work was carried out.

\appendix
\section{\label{app:1} Adiabatic analysis of Eqs.~(\ref{a-PT})}
In this appendix, we analyze the perturbation equations~\eqref{a-PT} for time-dependent coefficients $a_{\vq,n}(t)$. In the low-frequency limit [i.e., under assumption~\eqref{cond1}], we analyze the temporal integral in Eq.~\eqref{a-rec} separating the contribution of the upper limit $t'=t$ [we denote this contribution by the superscript ``0'': $a_{\vq,n}^0$] and the saddle-point vicinities of the integrand phase [we denote this contribution by the superscript ``\sp'': $a_{\vq,n}^\sp$]. 

\subsection{\label{app:A1} The contribution of the point $t'=t$}

The contribution of the point $t'=t$ leads to an account of the instantaneous ``interaction'' $\hat{\cV}(t)$ perturbatively. For convenience, we extract the dynamic phase \textcolor{violet}{$-i\int^t\cE(\tau)d\tau$} (where the function $\cE(t)$ is to be defined) from all coefficients $a^{0}_{\vq,n}(t)$, introducing $\alpha_{n}(t)$ (we omit the index $\vq$ for simplicity):
\begin{equation}
\label{a-alpha}
a^{0}_{\vq,n}(t) = \alpha_{n}(t)e^{-i\int^t \cE(\tau)d\tau}.
\end{equation}
The function $\cE(t)$ involves the influence of the interband couplings $\hat{\cV}(t)$ and tends to energy $E_0(t)$ of the electron in the initial band at $\hat{\cV}(t)\to 0$. We assume, that the exponential factor in Eq.~\eqref{a-alpha} sufficiently determines the time-dependence of $a^{0}_{\vq,n}(t)$, while the pre-exponential function $\alpha_{n}(t)$ is smooth and weakly depends on time:
\begin{equation}
\label{cond:alpha}
|\dot{\alpha}_n(t)|\ll|\alpha_n(t)\cE(t)|.
\end{equation}
This assumption is in agreement with the lowest adiabatic approximation and equivalent to consider only the contribution of the point $t'=t$ in Eq.~\eqref{a-rec}.

With the condition~\eqref{cond:alpha}, we obtain the approximate equation for $\alpha_n(t)$ from Eq.~\eqref{a(t)-Eq}:
\begin{equation}
\label{alpha:Eq}
[\cE(t)-E_n(t)]\alpha_n(t) = \sum_{n'}\cV_{nn'}(t)\alpha_{n'}(t).
\end{equation}
In Eq.~\eqref{alpha:Eq}, all quantities depend on $t$ parametrically, so that this equation can be analyzed within the standard approach of stationary perturbation theory (see, e.g.,~\cite{LL3}). Representing the energy $\cE(t)$ and coefficients $\alpha_n(t)$ by the perturbation series:
\begin{subequations}
\label{series:E-alpha}
\begin{eqnarray}
&& \cE(t) \equiv \cE_0(t) = E_0(t) + \sum_{m=1}^{\infty}\cE_0^{(m)}(t),  \\
&& \alpha_n(t) = \delta_{n0} + \sum_{m=1}^{\infty}\alpha_n^{(m)}(t),
\end{eqnarray}    
\end{subequations}
substituting Eqs.~\eqref{series:E-alpha} into Eq.~\eqref{alpha:Eq}, and equating the terms of the same perturbation order, for corrections $\cE_0^{(m)}(t)$, $\alpha_n^{(m)}$ ($\sim\hat{\cV}^m$), we obtain: 
\begin{subequations}
\label{corrections:E-alpha}
\begin{eqnarray}
\label{corr-E}
&& \cE_0^{(m)}(t) = \langle u_{\vQ(t),0}|\hat{\cV}(t)\big[G'_{E_0(t)}\hat\cV(t)\big]^{m-1} |u_{\vQ(t),0}\rangle,\\
\label{corr-alpha}
&& \alpha_n^{(m)}(t) = \frac{\langle u_{\vQ(t),n}|\hat{\cV}(t)\big[G'_{E_0(t)}\hat\cV(t)\big]^{m-1} |u_{\vQ(t),0}\rangle}{E_0(t)-E_n(t)}.
\;\;\;\;\;\;\;\;\;\;\;
\end{eqnarray}
\end{subequations}
In Eqs.~\eqref{corrections:E-alpha}, the reduced Green's function $G_{E_0(t)}'\equiv G_{E_0(t)}'(\vr,\vr')$ is used [see Eq.~\eqref{Green-red}].

Using expressions~\eqref{corrections:E-alpha}, we present the results for $\cE_0(t)$ and $a^{0}_{\vq,n}(t)$ in the form:
\begin{eqnarray}
\label{app:cE}
 \cE_0(t) &=& E_0(t)+\Delta E_0(t),\\  
\label{app:Delta_cE}
 \Delta E_0(t)&=&\langle u_{\vQ(\tau),0}|\hat{\cV}(\tau)|\cU_{\vQ(\tau),0}\rangle,\\
\label{0:a:fin}
\nonumber
 a_{\vq,n}^0(t)&=& e^{-i\int^t \cE_0(\tau)d\tau}\Big[
\delta_{n0} \\ 
 \qquad && + (1-\delta_{n0})\frac{\langle u_{\vQ(t),n}|\hat{\cV}(t)|\cU_{\vQ(t),0}\rangle}{E_0(t)-E_n(t)}
\Bigg],
\end{eqnarray}
where
\begin{eqnarray}
\label{app:cU}
\nonumber
|\cU_{\vQ(t),0}\rangle &=& \sum_{m=0}^{\infty}\big[G'_{E_0(t)}\hat\cV(t)\big]^m |u_{\vQ(t),0}\rangle \\
\qquad &=& |u_{\vQ(t),0}\rangle + G'_{E_0(t)}\hat\cV(t)|\cU_{\vQ(t),0}\rangle.
\end{eqnarray}

We assume convergence of the perturbation series in Eq.~\eqref{series:E-alpha}. The rigorous justification of this assumption deserves separate consideration.

\begin{widetext}

\subsection{The saddle-point contribution}

We represent the coefficients $a_{\vq,n}^{(m)}(t)$ in the following form:
\begin{eqnarray}
\label{app:a-recur}
\nonumber
&&a^{(m)}_{\vq,n}(t) = -i\int^t_{-\infty}dt_m\int^{t_{m}}_{-\infty}dt_{m-1} \cdots \int^{t_{2}}_{-\infty}dt_1 
e^{-i\left[\int_{t_m}^tE_n(\tau)d\tau+\int^{t_1}E_0(\tau)d\tau\right]}\\
&&\phantom{a^{(m)}_{\vq,n}(t) =}\times \langle u_{\vQ(t_m),n}|\hat{\cV}(t_m) G(t_m,t_{m-1})\hat\cV(t_{m-1})\cdots G(t_{2},t_1)\hat{\cV}(t_1)|
u_{\vQ(t_1),0}\rangle,
\end{eqnarray}    
where $G(t,t')\equiv G(\vr,t;\vr',t')$ is the non-stationary Green's function on the basis of the Houston functions [cf.~\eqref{Houston-psi}], 
\begin{equation}
\label{Green-q}
G(t,t') = -i\sum_{n}e^{-i\int^t_{t'}E_n(\tau)d\tau}|u_{\vQ(t),n}\rangle\langle u_{\vQ(t'),n}|.
\end{equation}
Estimating the saddle-point contribution to the integrals in Eq.~\eqref{app:a-recur}, it should be taken into account that this contribution is defined by the exponentially small tunneling factor. For this reason, in the lowest order, we limit ourselves to applying the saddle-point estimation only to a single integral in Eq.~\eqref{app:a-recur} over one of the times $t_k$, $k=1,...,m$ (we denote the corresponding intergral by the superscript ``sp''), while all other integrals should be estimated taking into account only the upper-limits contributions:
\begin{eqnarray}
\label{app:a-recur-sp}
\nonumber
&&a^{(m)}_{\vq,n}(t) = (-i)^3\sum_{k=1}^m\sum_{\nu,\nu'}\int^t_{-\infty}dt_m\cdots\int^{t_{k+2}}_{-\infty}dt_{k+1} 
e^{-i\int_{t_m}^tE_n(\tau)d\tau}\\
\nonumber
&&\phantom{a^{(m)}_{\vq,n}(t) =}\times \langle u_{\vQ(t_m),n}|\hat{\cV}(t_m) G(t_m,t_{m-1})\cdots G(t_{k+2},t_{k+1})\hat{\cV}(t_{k+1})|u_{\vQ(t_{k+1}),\nu}\rangle e^{-i\int^{t_{k+1}}E_{\nu}(\tau)d\tau}\\
\nonumber
&&\phantom{a^{(m)}_{\vq,n}(t) =}\times \int^{(\sp)} dt_{k}\; e^{i\int^{t_k} [E_{\nu}(\tau)-E_{\nu'}(\tau)]d\tau}\cV_{\nu\nu'}(t_k)
\int^{t_{k}}_{-\infty}dt_{k-1} \cdots \int^{t_{2}}_{-\infty}dt_1 \\
&&\phantom{a^{(m)}_{\vq,n}(t) =}\times e^{i\int^{t_{k-1}}E_{\nu'}(\tau)d\tau}\langle u_{\vQ(t_{k-1}),\nu'}|\hat{\cV}(t_{k-1})G(t_{k-1},t_{k-2})\cdots G(t_{2},t_1)\hat{\cV}(t_1)|
u_{\vQ(t_1),0}\rangle e^{-i\int^{t_1}E_0(\tau)d\tau}.
\end{eqnarray} 
In Eq.~\eqref{app:a-recur-sp}, integrals over times $t_1$, $t_2\dots t_{k-1}$ lead to the result for $\alpha_{\nu'}^{(k-1)}(t_k)$ according to the analysis in Sec.~\ref{app:A1} (only the contribution of the vicinity of $t_1=t_2\cdots=t_{k-1}= t_k$ is taken into account):
\begin{eqnarray}
\label{app:alpha_nu1}
\nonumber
&& \int^{t_{k}}_{-\infty}dt_{k-1} \cdots \int^{t_{2}}_{-\infty}dt_1 e^{i\int^{t_{k-1}}E_{\nu'}(\tau)d\tau}\langle u_{\vQ(t_{k-1}),\nu'}|\hat{\cV}(t_{k-1})G(t_{k-1},t_{k-2})\cdots G(t_{2},t_1)\hat{\cV}(t_1)|
u_{\vQ(t_1),0}\rangle e^{-i\int^{t_1}E_0(\tau)d\tau} \\
&&  = \alpha_{\nu'}^{(k-1)}(t_k)e^{i\int^{t_k}\left[E_{\nu'}(\tau)-\cE_0(\tau)\right]d\tau},
\end{eqnarray} 
while within the same approximation (i.e., considering the contribution of the vicinity of $t_{k+1}=t_{k+2}\cdots=t_{m}= t$), the integrals over times $t_{k+1}$, $t_{k+2}\dots t_{m}$ give the following result:
\begin{eqnarray}
\label{app:alpha_nu}
\nonumber
&& \int^{t}_{-\infty}dt_{m} \cdots \int^{t_{k+2}}_{-\infty}dt_{k+1} e^{-i\int_{t_m}^tE_n(\tau)d\tau}\langle u_{\vQ(t_m),n}|\hat{\cV}(t_m) G(t_m,t_{m-1})\cdots G(t_{k+2},t_{k+1})\hat{\cV}(t_{k+1})|u_{\vQ(t_{k+1}),\nu}\rangle \\ 
&& \times e^{-i\int^{t_{k+1}}E_{\nu}(\tau)d\tau} = \alpha_{n(\nu)}^{(m-k)}(t)e^{-i\int^{t}\cE_{\nu}(\tau)d\tau},
\end{eqnarray} 
where $\alpha_{n(\nu)}^{(m-k)}(t)$ is determined by Eq.~\eqref{corr-alpha} by replacing there the index $0$ by $\nu$.

Taking into account Eqs.~\eqref{app:alpha_nu1}, \eqref{app:alpha_nu} and evaluating the sum over all perturbation orders $m$ in accordance with Eq.~\eqref{0:a:fin}, from Eq.~\eqref{app:a-recur-sp}, we arrive at the following expression:
\begin{eqnarray}
\label{sp:a-int}
\nonumber
&& a^{\sp}_{\vq,n}(t) = \sum_{m=1}^{\infty}a^{(m)}_{\vq,n}(t)
= i\sum_{\nu,\nu'} a^{0}_{\vq,n(\nu)}(t)\int^{(\sp)} e^{i\int^{t'}E_\nu(\tau)d\tau}\cV_{\nu\nu'}(t')a^{0}_{\vq,\nu'}(t')dt'\\
&& \phantom{a^{\sp}_{\vq,n}(t)} = i\sum_{\nu} a^{0}_{\vq,n(\nu)}(t)\int^{(\sp)} e^{i\int^{t'}\left[E_\nu(\tau)-\cE_0(\tau)\right]d\tau}\langle u_{\vQ(t'),\nu}|\hat{\cV}|\cU_{\vQ(t'),0}\rangle dt',
\end{eqnarray}
\end{widetext}
where $a^{0}_{\vq,n(\nu)}(t)$ are defined by Eq.~\eqref{0:a:fin}, in which the index $0$ should be replaced by $\nu$.

Within condition~\eqref{cond2}, for estimation of the integral in Eq.~\eqref{sp:a-int}, we can neglect the effect of the band-coupling correction $\Delta E_0$ in Eq.~\eqref{app:cE} on the saddle points, considering the exponential $\exp[-i\int^{t'}\Delta E_0(\tau)d\tau]$ to be a slow-varying function of $t'$. Then the saddle points $t'_s{}^{(\sp)}$ satisfy the equation:
\begin{equation}
\label{sp-Eq}
\Delta_{\nu 0}(t'_s{}^{(\sp)}) = E_\nu(t'_s{}^{(\sp)}) - E_0(t'_s{}^{(\sp)}) =0.
\end{equation}
Due to the non-zero band gap $\Delta_{\nu0}(t')$ for real $t'$, the solutions of Eq.~\eqref{sp-Eq} are complex: $t'_s{}^{(\sp)}=t'_s + i\tau_s$. We assume that the imaginary parts $\tau_s$ are small ($\tau_s\ll t'_s$) and can be taken into account perturbatively. Expanding the terms in Eq.~\eqref{sp-Eq} in $\tau_s$ up to the second power $\sim\tau_s^2$, we obtain:
\begin{subequations}
\begin{eqnarray}
\label{speq-1}
&&\Delta_{\nu0}(t_s') = \frac{\partial^2\Delta_{\nu0}(t'_s)}{\partial t'^2_s}\frac{\tau_s^2}{2},\\
\label{speq-2}
&&\frac{\partial\Delta_{\nu0}(t'_s)}{\partial t'_s} = 0.
\end{eqnarray}
\end{subequations}
The result for $\tau_s$ immediately follows from Eq.~\eqref{speq-1}: 
\begin{subequations}
\begin{eqnarray}
\label{tau_s}
&&\tau_s = \pm\frac{\varkappa_{\nu0}}{\cF_{\nu0}},\\
\label{app:varkappa}
&&\varkappa_{\nu0} \equiv \varkappa_{\nu0}(t'_s) = \sqrt{2\Delta_{\nu0}(t_s')},\\ 
\label{app:calF}
\nonumber
&&\cF_{\nu0} \equiv \cF_{\nu0}(t_s') \\
&& = \sqrt{\vF^2(t'_s)\nabla_\vq^2\Delta_{\nu0}(t_s') - \dot{\vF}(t'_s)\nabla_\vq\Delta_{\nu0}(t_s')},\;\;\;\;\;\;\;
\end{eqnarray} 
\end{subequations}
and the real times $t'_s$ satisfy Eq.~\eqref{speq-2} or
\begin{equation}
\label{app:speq-real}
{\vF}(t'_s)\cdot\nabla_\vq\Delta_{\nu0}(t_s') = 0.
\end{equation} 

Evaluating the imaginary part of the phase $\int^{t'_s{}^{(\sp)}}_{\infty}\Delta_{\nu0}(\tau)d\tau$ [$\Delta_{\nu0}(\tau)=E_\nu(\tau)-E_0(\tau)$] in Eq.~\eqref{sp:a-int} at $t'_s{}^{(\sp)}=t'_s+i\tau_s$ as 
\begin{equation*}
{\rm Im}\int^{t'_s{}^{(\sp)}}_{\infty}\Delta_{\nu0}(\tau)d\tau = - \frac{\partial^2\Delta_{\nu0}(t_s')}{\partial t_s'^2}\frac{\tau_s^3}{6} = 
\mp \frac{\varkappa_{\nu0}^3}{3\cF_{\nu0}},
\end{equation*}
we conclude that the positive value $\tau_s$ leads to an exponential increase of $a^{\sp}_{\vq,n}$ with increasing band gap $\Delta_{\nu 0}$ or vanishing the effective field strength $\cF_{\nu 0}$. For this reason, we should leave the negative solution in Eq.~\eqref{tau_s}, while the saddle-point result for $a^{\sp}_{\vq,n}(t)$ is:
\begin{eqnarray}
\label{sp:a:fin}
\nonumber
&&a^{\sp}_{\vq,n}(t) = \sqrt{\frac{2\pi}{i}}\sum_{\nu} a^{0}_{\vq,n(\nu)}(t)\sum_{s}\frac{e^{-\frac{\varkappa^3_{\nu 0}}{3\cF_{\nu 0}}}}{\sqrt{\cF_{\nu 0}\varkappa_{\nu 0}}} \\
&&\quad \times \langle u_{\vQ(t_s'),\nu}|\hat{\cV}|\cU_{\vQ(t_s'),0}\rangle e^{i\int^{t'_s}\left[E_\nu(\tau)-\cE_0(\tau)\right]d\tau}.\;\;\;\;
\end{eqnarray}


\begin{thebibliography}{38}%
\makeatletter
\providecommand \@ifxundefined [1]{%
 \@ifx{#1\undefined}
}%
\providecommand \@ifnum [1]{%
 \ifnum #1\expandafter \@firstoftwo
 \else \expandafter \@secondoftwo
 \fi
}%
\providecommand \@ifx [1]{%
 \ifx #1\expandafter \@firstoftwo
 \else \expandafter \@secondoftwo
 \fi
}%
\providecommand \natexlab [1]{#1}%
\providecommand \enquote  [1]{``#1''}%
\providecommand \bibnamefont  [1]{#1}%
\providecommand \bibfnamefont [1]{#1}%
\providecommand \citenamefont [1]{#1}%
\providecommand \href@noop [0]{\@secondoftwo}%
\providecommand \href [0]{\begingroup \@sanitize@url \@href}%
\providecommand \@href[1]{\@@startlink{#1}\@@href}%
\providecommand \@@href[1]{\endgroup#1\@@endlink}%
\providecommand \@sanitize@url [0]{\catcode `\\12\catcode `\$12\catcode
  `\&12\catcode `\#12\catcode `\^12\catcode `\_12\catcode `\%12\relax}%
\providecommand \@@startlink[1]{}%
\providecommand \@@endlink[0]{}%
\providecommand \url  [0]{\begingroup\@sanitize@url \@url }%
\providecommand \@url [1]{\endgroup\@href {#1}{\urlprefix }}%
\providecommand \urlprefix  [0]{URL }%
\providecommand \Eprint [0]{\href }%
\providecommand \doibase [0]{http://dx.doi.org/}%
\providecommand \selectlanguage [0]{\@gobble}%
\providecommand \bibinfo  [0]{\@secondoftwo}%
\providecommand \bibfield  [0]{\@secondoftwo}%
\providecommand \translation [1]{[#1]}%
\providecommand \BibitemOpen [0]{}%
\providecommand \bibitemStop [0]{}%
\providecommand \bibitemNoStop [0]{.\EOS\space}%
\providecommand \EOS [0]{\spacefactor3000\relax}%
\providecommand \BibitemShut  [1]{\csname bibitem#1\endcsname}%
\let\auto@bib@innerbib\@empty
\bibitem [{\citenamefont {Ferray}\ \emph {et~al.}(1988)\citenamefont {Ferray},
  \citenamefont {L'Huillier}, \citenamefont {Li}, \citenamefont {Lompre},
  \citenamefont {Mainfray},\ and\ \citenamefont {Manus}}]{LHuillier88}%
  \BibitemOpen
  \bibfield  {author} {\bibinfo {author} {\bibfnamefont {M.}~\bibnamefont
  {Ferray}}, \bibinfo {author} {\bibfnamefont {A.}~\bibnamefont {L'Huillier}},
  \bibinfo {author} {\bibfnamefont {X.~F.}\ \bibnamefont {Li}}, \bibinfo
  {author} {\bibfnamefont {L.~A.}\ \bibnamefont {Lompre}}, \bibinfo {author}
  {\bibfnamefont {G.}~\bibnamefont {Mainfray}}, \ and\ \bibinfo {author}
  {\bibfnamefont {C.}~\bibnamefont {Manus}},\ }\bibfield  {title} {\enquote
  {\bibinfo {title} {Multiple-harmonic conversion of 1064 nm radiation in rare
  gases},}\ }\href@noop {} {\bibfield  {journal} {\bibinfo  {journal} {J. Phys.
  B: At. Mol. Opt. Phys.}\ }\textbf {\bibinfo {volume} {21}},\ \bibinfo {pages}
  {L31} (\bibinfo {year} {1988})}\BibitemShut {NoStop}%
\bibitem [{\citenamefont {L'Huillier}\ \emph {et~al.}(1991)\citenamefont
  {L'Huillier}, \citenamefont {Schafer},\ and\ \citenamefont
  {Kulander}}]{HSKJPB91}%
  \BibitemOpen
  \bibfield  {author} {\bibinfo {author} {\bibfnamefont {A.}~\bibnamefont
  {L'Huillier}}, \bibinfo {author} {\bibfnamefont {K.~J.}\ \bibnamefont
  {Schafer}}, \ and\ \bibinfo {author} {\bibfnamefont {K.~C.}\ \bibnamefont
  {Kulander}},\ }\bibfield  {title} {\enquote {\bibinfo {title} {Theoretical
  aspects of intense field harmonic generation},}\ }\href@noop {} {\bibfield
  {journal} {\bibinfo  {journal} {J. Phys. B: At. Mol. Opt. Phys.}\ }\textbf
  {\bibinfo {volume} {24}},\ \bibinfo {pages} {3315--3341} (\bibinfo {year}
  {1991})}\BibitemShut {NoStop}%
\bibitem [{\citenamefont {Scrinzi}\ \emph {et~al.}(2006)\citenamefont
  {Scrinzi}, \citenamefont {Ivanov}, \citenamefont {Kienberger},\ and\
  \citenamefont {Villeneuve}}]{SIKVJPB06}%
  \BibitemOpen
  \bibfield  {author} {\bibinfo {author} {\bibfnamefont {A}~\bibnamefont
  {Scrinzi}}, \bibinfo {author} {\bibfnamefont {M.~Yu.}\ \bibnamefont
  {Ivanov}}, \bibinfo {author} {\bibfnamefont {R.}~\bibnamefont {Kienberger}},
  \ and\ \bibinfo {author} {\bibfnamefont {D.~M.}\ \bibnamefont {Villeneuve}},\
  }\bibfield  {title} {\enquote {\bibinfo {title} {Attosecond physics},}\
  }\href@noop {} {\bibfield  {journal} {\bibinfo  {journal} {J. Phys. B: At.
  Mol. Opt. Phys.}\ }\textbf {\bibinfo {volume} {39}},\ \bibinfo {pages}
  {R1--R37} (\bibinfo {year} {2006})}\BibitemShut {NoStop}%
\bibitem [{\citenamefont {Krausz}\ and\ \citenamefont
  {Ivanov}(2009)}]{KIRMP09}%
  \BibitemOpen
  \bibfield  {author} {\bibinfo {author} {\bibfnamefont {F.}~\bibnamefont
  {Krausz}}\ and\ \bibinfo {author} {\bibfnamefont {M.}~\bibnamefont
  {Ivanov}},\ }\bibfield  {title} {\enquote {\bibinfo {title} {Attosecond
  physics},}\ }\href {\doibase 10.1103/RevModPhys.81.163} {\bibfield  {journal}
  {\bibinfo  {journal} {Rev. Mod. Phys.}\ }\textbf {\bibinfo {volume} {81}},\
  \bibinfo {pages} {163} (\bibinfo {year} {2009})}\BibitemShut {NoStop}%
\bibitem [{\citenamefont {Schafer}\ \emph {et~al.}(1993)\citenamefont
  {Schafer}, \citenamefont {Yang}, \citenamefont {DiMauro},\ and\ \citenamefont
  {Kulander}}]{SYDKPRL93}%
  \BibitemOpen
  \bibfield  {author} {\bibinfo {author} {\bibfnamefont {K.~J.}\ \bibnamefont
  {Schafer}}, \bibinfo {author} {\bibfnamefont {Baorui}\ \bibnamefont {Yang}},
  \bibinfo {author} {\bibfnamefont {L.~F.}\ \bibnamefont {DiMauro}}, \ and\
  \bibinfo {author} {\bibfnamefont {K.~C.}\ \bibnamefont {Kulander}},\
  }\bibfield  {title} {\enquote {\bibinfo {title} {Above threshold ionization
  beyond the high harmonic cutoff},}\ }\href@noop {} {\bibfield  {journal}
  {\bibinfo  {journal} {Phys. Rev. Lett.}\ }\textbf {\bibinfo {volume} {70}},\
  \bibinfo {pages} {1599--1602} (\bibinfo {year} {1993})}\BibitemShut {NoStop}%
\bibitem [{\citenamefont {Corkum}(1993)}]{CorkumPRL93}%
  \BibitemOpen
  \bibfield  {author} {\bibinfo {author} {\bibfnamefont {P.~B.}\ \bibnamefont
  {Corkum}},\ }\bibfield  {title} {\enquote {\bibinfo {title} {Plasma
  perspective on strong field multiphoton ionization},}\ }\href {\doibase
  10.1103/PhysRevLett.71.1994} {\bibfield  {journal} {\bibinfo  {journal}
  {Phys. Rev. Lett.}\ }\textbf {\bibinfo {volume} {71}},\ \bibinfo {pages}
  {1994} (\bibinfo {year} {1993})}\BibitemShut {NoStop}%
\bibitem [{\citenamefont {Paulus}\ \emph {et~al.}(1994)\citenamefont {Paulus},
  \citenamefont {Becker}, \citenamefont {Nicklich},\ and\ \citenamefont
  {Walther}}]{PBNWJPB94}%
  \BibitemOpen
  \bibfield  {author} {\bibinfo {author} {\bibfnamefont {G~G}\ \bibnamefont
  {Paulus}}, \bibinfo {author} {\bibfnamefont {W}~\bibnamefont {Becker}},
  \bibinfo {author} {\bibfnamefont {W}~\bibnamefont {Nicklich}}, \ and\
  \bibinfo {author} {\bibfnamefont {H}~\bibnamefont {Walther}},\ }\bibfield
  {title} {\enquote {\bibinfo {title} {Rescattering effects in above-threshold
  ionization: a classical model},}\ }\href {\doibase
  https://doi.org/10.1088/0953-4075/27/21/003} {\bibfield  {journal} {\bibinfo
  {journal} {J. Phys. B: At. Mol. Opt. Phys.}\ }\textbf {\bibinfo {volume}
  {27}},\ \bibinfo {pages} {L703} (\bibinfo {year} {1994})}\BibitemShut
  {NoStop}%
\bibitem [{\citenamefont {Ghimire}\ \emph {et~al.}(2011)\citenamefont
  {Ghimire}, \citenamefont {DiChiara}, \citenamefont {Sistrunk}, \citenamefont
  {Agostini}, \citenamefont {DiMauro},\ and\ \citenamefont
  {Reis}}]{Ghimire2011}%
  \BibitemOpen
  \bibfield  {author} {\bibinfo {author} {\bibfnamefont {S.}~\bibnamefont
  {Ghimire}}, \bibinfo {author} {\bibfnamefont {A.~D.}\ \bibnamefont
  {DiChiara}}, \bibinfo {author} {\bibfnamefont {E.}~\bibnamefont {Sistrunk}},
  \bibinfo {author} {\bibfnamefont {P.}~\bibnamefont {Agostini}}, \bibinfo
  {author} {\bibfnamefont {L.~F.}\ \bibnamefont {DiMauro}}, \ and\ \bibinfo
  {author} {\bibfnamefont {D.~A.}\ \bibnamefont {Reis}},\ }\bibfield  {title}
  {\enquote {\bibinfo {title} {Observation of high-order harmonic generation in
  a bulk crystal},}\ }\href {\doibase 10.1038/nphys1847} {\bibfield  {journal}
  {\bibinfo  {journal} {Nature Physics}\ }\textbf {\bibinfo {volume} {7}},\
  \bibinfo {pages} {138--141} (\bibinfo {year} {2011})}\BibitemShut {NoStop}%
\bibitem [{\citenamefont {Kruchinin}\ \emph {et~al.}(2018)\citenamefont
  {Kruchinin}, \citenamefont {Krausz},\ and\ \citenamefont
  {Yakovlev}}]{KKYRMP18}%
  \BibitemOpen
  \bibfield  {author} {\bibinfo {author} {\bibfnamefont {S.~Yu.}\ \bibnamefont
  {Kruchinin}}, \bibinfo {author} {\bibfnamefont {F.}~\bibnamefont {Krausz}}, \
  and\ \bibinfo {author} {\bibfnamefont {V.~S.}\ \bibnamefont {Yakovlev}},\
  }\bibfield  {title} {\enquote {\bibinfo {title} {Colloquium: Strong-field
  phenomena in periodic systems},}\ }\href {\doibase
  10.1103/RevModPhys.90.021002} {\bibfield  {journal} {\bibinfo  {journal}
  {Rev. Mod. Phys.}\ }\textbf {\bibinfo {volume} {90}},\ \bibinfo {pages}
  {021002} (\bibinfo {year} {2018})}\BibitemShut {NoStop}%
\bibitem [{\citenamefont {Park}\ \emph {et~al.}(2022)\citenamefont {Park},
  \citenamefont {Subramani}, \citenamefont {Kim},\ and\ \citenamefont
  {Ciappina}}]{Park22}%
  \BibitemOpen
  \bibfield  {author} {\bibinfo {author} {\bibfnamefont {Jongkyoon}\
  \bibnamefont {Park}}, \bibinfo {author} {\bibfnamefont {Amutha}\ \bibnamefont
  {Subramani}}, \bibinfo {author} {\bibfnamefont {Seungchul}\ \bibnamefont
  {Kim}}, \ and\ \bibinfo {author} {\bibfnamefont {Marcelo~F.}\ \bibnamefont
  {Ciappina}},\ }\bibfield  {title} {\enquote {\bibinfo {title} {Recent trends
  in high-order harmonic generation in solids},}\ }\href {\doibase
  10.1080/23746149.2021.2003244} {\bibfield  {journal} {\bibinfo  {journal}
  {Advances in Physics: X}\ }\textbf {\bibinfo {volume} {7}},\ \bibinfo {pages}
  {2003244} (\bibinfo {year} {2022})}\BibitemShut {NoStop}%
\bibitem [{\citenamefont {Goulielmakis}\ and\ \citenamefont
  {Brabec}(2022)}]{Goulielmakis2022}%
  \BibitemOpen
  \bibfield  {author} {\bibinfo {author} {\bibfnamefont {Eleftherios}\
  \bibnamefont {Goulielmakis}}\ and\ \bibinfo {author} {\bibfnamefont {Thomas}\
  \bibnamefont {Brabec}},\ }\bibfield  {title} {\enquote {\bibinfo {title}
  {High harmonic generation in condensed matter},}\ }\href {\doibase
  10.1038/s41566-022-00988-y} {\bibfield  {journal} {\bibinfo  {journal} {Nat.
  Photon.}\ }\textbf {\bibinfo {volume} {16}},\ \bibinfo {pages} {411}
  (\bibinfo {year} {2022})}\BibitemShut {NoStop}%
\bibitem [{\citenamefont {Yue}\ and\ \citenamefont {Gaarde}(2022)}]{Yue2022}%
  \BibitemOpen
  \bibfield  {author} {\bibinfo {author} {\bibfnamefont {L.}~\bibnamefont
  {Yue}}\ and\ \bibinfo {author} {\bibfnamefont {M.~B.}\ \bibnamefont
  {Gaarde}},\ }\bibfield  {title} {\enquote {\bibinfo {title} {Introduction to
  theory of high-harmonic generation in solids: tutorial},}\ }\href {\doibase
  10.1364/josab.448602} {\bibfield  {journal} {\bibinfo  {journal} {JOSA B}\
  }\textbf {\bibinfo {volume} {39}},\ \bibinfo {pages} {535} (\bibinfo {year}
  {2022})}\BibitemShut {NoStop}%
\bibitem [{\citenamefont {Ghimire}\ and\ \citenamefont
  {Reis}(2019)}]{Ghimire2019}%
  \BibitemOpen
  \bibfield  {author} {\bibinfo {author} {\bibfnamefont {S.}~\bibnamefont
  {Ghimire}}\ and\ \bibinfo {author} {\bibfnamefont {D.~A.}\ \bibnamefont
  {Reis}},\ }\bibfield  {title} {\enquote {\bibinfo {title} {High-harmonic
  generation from solids},}\ }\href {\doibase 10.1038/s41567-018-0315-5}
  {\bibfield  {journal} {\bibinfo  {journal} {Nature Physics}\ }\textbf
  {\bibinfo {volume} {15}},\ \bibinfo {pages} {10--16} (\bibinfo {year}
  {2019})}\BibitemShut {NoStop}%
\bibitem [{\citenamefont {Li}\ \emph {et~al.}(2023)\citenamefont {Li},
  \citenamefont {Lan}, \citenamefont {Zhu},\ and\ \citenamefont
  {Lu}}]{LiRPP2023}%
  \BibitemOpen
  \bibfield  {author} {\bibinfo {author} {\bibfnamefont {L.}~\bibnamefont
  {Li}}, \bibinfo {author} {\bibfnamefont {P.}~\bibnamefont {Lan}}, \bibinfo
  {author} {\bibfnamefont {X.}~\bibnamefont {Zhu}}, \ and\ \bibinfo {author}
  {\bibfnamefont {P.}~\bibnamefont {Lu}},\ }\bibfield  {title} {\enquote
  {\bibinfo {title} {High harmonic generation in solids: particle and wave
  perspectives},}\ }\href {\doibase 10.1088/1361-6633/acf144} {\bibfield
  {journal} {\bibinfo  {journal} {Rep. Prog. Phys.}\ }\textbf {\bibinfo
  {volume} {86}},\ \bibinfo {pages} {116401} (\bibinfo {year}
  {2023})}\BibitemShut {NoStop}%
\bibitem [{\citenamefont {Ghimire}\ \emph {et~al.}(2012)\citenamefont
  {Ghimire}, \citenamefont {DiChiara}, \citenamefont {Sistrunk}, \citenamefont
  {Ndabashimiye}, \citenamefont {Szafruga}, \citenamefont {Mohammad},
  \citenamefont {Agostini}, \citenamefont {DiMauro},\ and\ \citenamefont
  {Reis}}]{GhimirePRA12}%
  \BibitemOpen
  \bibfield  {author} {\bibinfo {author} {\bibfnamefont {S.}~\bibnamefont
  {Ghimire}}, \bibinfo {author} {\bibfnamefont {A.~D.}\ \bibnamefont
  {DiChiara}}, \bibinfo {author} {\bibfnamefont {E.}~\bibnamefont {Sistrunk}},
  \bibinfo {author} {\bibfnamefont {G.}~\bibnamefont {Ndabashimiye}}, \bibinfo
  {author} {\bibfnamefont {U.~B.}\ \bibnamefont {Szafruga}}, \bibinfo {author}
  {\bibfnamefont {A.}~\bibnamefont {Mohammad}}, \bibinfo {author}
  {\bibfnamefont {P.}~\bibnamefont {Agostini}}, \bibinfo {author}
  {\bibfnamefont {L.~F.}\ \bibnamefont {DiMauro}}, \ and\ \bibinfo {author}
  {\bibfnamefont {D.~A.}\ \bibnamefont {Reis}},\ }\bibfield  {title} {\enquote
  {\bibinfo {title} {Generation and propagation of high-order harmonics in
  crystals},}\ }\href {\doibase 10.1103/PhysRevA.85.043836} {\bibfield
  {journal} {\bibinfo  {journal} {PRA}\ }\textbf {\bibinfo {volume} {85}},\
  \bibinfo {pages} {043836} (\bibinfo {year} {2012})}\BibitemShut {NoStop}%
\bibitem [{\citenamefont {Vampa}\ \emph {et~al.}(2014)\citenamefont {Vampa},
  \citenamefont {McDonald}, \citenamefont {Orlando}, \citenamefont {Klug},
  \citenamefont {Corkum},\ and\ \citenamefont {Brabec}}]{VampaPRL14}%
  \BibitemOpen
  \bibfield  {author} {\bibinfo {author} {\bibfnamefont {G.}~\bibnamefont
  {Vampa}}, \bibinfo {author} {\bibfnamefont {C.~R.}\ \bibnamefont {McDonald}},
  \bibinfo {author} {\bibfnamefont {G.}~\bibnamefont {Orlando}}, \bibinfo
  {author} {\bibfnamefont {D.~D.}\ \bibnamefont {Klug}}, \bibinfo {author}
  {\bibfnamefont {P.~B.}\ \bibnamefont {Corkum}}, \ and\ \bibinfo {author}
  {\bibfnamefont {T.}~\bibnamefont {Brabec}},\ }\bibfield  {title} {\enquote
  {\bibinfo {title} {Theoretical analysis of high-harmonic generation in
  solids},}\ }\href {\doibase 10.1103/PhysRevLett.113.073901} {\bibfield
  {journal} {\bibinfo  {journal} {Phys. Rev. Lett.}\ }\textbf {\bibinfo
  {volume} {113}},\ \bibinfo {pages} {073901} (\bibinfo {year}
  {2014})}\BibitemShut {NoStop}%
\bibitem [{\citenamefont {Vampa}\ and\ \citenamefont
  {Brabec}(2015)}]{Vampa2015}%
  \BibitemOpen
  \bibfield  {author} {\bibinfo {author} {\bibfnamefont {G.}~\bibnamefont
  {Vampa}}\ and\ \bibinfo {author} {\bibfnamefont {T.}~\bibnamefont {Brabec}},\
  }\bibfield  {title} {\enquote {\bibinfo {title} {Semiclassical analysis of
  high harmonic generation in bulk crystals},}\ }\href {\doibase
  10.1103/PhysRevB.91.064302} {\bibfield  {journal} {\bibinfo  {journal} {J.
  Phys. B: At. Mol. Opt. Phys.}\ }\textbf {\bibinfo {volume} {91}},\ \bibinfo
  {pages} {064302} (\bibinfo {year} {2015})}\BibitemShut {NoStop}%
\bibitem [{\citenamefont {Li}\ \emph {et~al.}(2019)\citenamefont {Li},
  \citenamefont {Lan}, \citenamefont {Zhu}, \citenamefont {Huang},
  \citenamefont {Zhang}, \citenamefont {Lein},\ and\ \citenamefont
  {Lu}}]{LiPRL19}%
  \BibitemOpen
  \bibfield  {author} {\bibinfo {author} {\bibfnamefont {L.}~\bibnamefont
  {Li}}, \bibinfo {author} {\bibfnamefont {P.}~\bibnamefont {Lan}}, \bibinfo
  {author} {\bibfnamefont {X.}~\bibnamefont {Zhu}}, \bibinfo {author}
  {\bibfnamefont {T.}~\bibnamefont {Huang}}, \bibinfo {author} {\bibfnamefont
  {Q.}~\bibnamefont {Zhang}}, \bibinfo {author} {\bibfnamefont
  {M.}~\bibnamefont {Lein}}, \ and\ \bibinfo {author} {\bibfnamefont
  {P.}~\bibnamefont {Lu}},\ }\bibfield  {title} {\enquote {\bibinfo {title}
  {Reciprocal-space-trajectory perspective on high-harmonic generation in
  solids},}\ }\href {\doibase 10.1103/PhysRevLett.122.193901} {\bibfield
  {journal} {\bibinfo  {journal} {Phys. Rev. Lett.}\ }\textbf {\bibinfo
  {volume} {122}},\ \bibinfo {pages} {193901} (\bibinfo {year}
  {2019})}\BibitemShut {NoStop}%
\bibitem [{\citenamefont {Golde}\ \emph {et~al.}(2008)\citenamefont {Golde},
  \citenamefont {Meier},\ and\ \citenamefont {Koch}}]{GoldePRB08}%
  \BibitemOpen
  \bibfield  {author} {\bibinfo {author} {\bibfnamefont {D.}~\bibnamefont
  {Golde}}, \bibinfo {author} {\bibfnamefont {T.}~\bibnamefont {Meier}}, \ and\
  \bibinfo {author} {\bibfnamefont {S.~W.}\ \bibnamefont {Koch}},\ }\bibfield
  {title} {\enquote {\bibinfo {title} {High harmonics generated in
  semiconductor nanostructures by the coupled dynamics of optical inter- and
  intraband excitations},}\ }\href {\doibase 10.1103/PhysRevB.77.075330}
  {\bibfield  {journal} {\bibinfo  {journal} {Phys. Rev. B}\ }\textbf {\bibinfo
  {volume} {77}},\ \bibinfo {pages} {075330} (\bibinfo {year}
  {2008})}\BibitemShut {NoStop}%
\bibitem [{\citenamefont {Luu}\ and\ \citenamefont
  {W\"orner}(2016)}]{LuuPRB16}%
  \BibitemOpen
  \bibfield  {author} {\bibinfo {author} {\bibfnamefont {T.~T.}\ \bibnamefont
  {Luu}}\ and\ \bibinfo {author} {\bibfnamefont {H.~J.}\ \bibnamefont
  {W\"orner}},\ }\bibfield  {title} {\enquote {\bibinfo {title} {High-order
  harmonic generation in solids: A unifying approach},}\ }\href {\doibase
  10.1103/PhysRevB.94.115164} {\bibfield  {journal} {\bibinfo  {journal} {Phys.
  Rev. B}\ }\textbf {\bibinfo {volume} {94}},\ \bibinfo {pages} {115164}
  (\bibinfo {year} {2016})}\BibitemShut {NoStop}%
\bibitem [{\citenamefont {Tancogne-Dejean}\ \emph {et~al.}(2017)\citenamefont
  {Tancogne-Dejean}, \citenamefont {M\"ucke}, \citenamefont {K\"artner},\ and\
  \citenamefont {Rubio}}]{TancognePRL17}%
  \BibitemOpen
  \bibfield  {author} {\bibinfo {author} {\bibfnamefont {N.}~\bibnamefont
  {Tancogne-Dejean}}, \bibinfo {author} {\bibfnamefont {O.~D.}\ \bibnamefont
  {M\"ucke}}, \bibinfo {author} {\bibfnamefont {F.~X.}\ \bibnamefont
  {K\"artner}}, \ and\ \bibinfo {author} {\bibfnamefont {A.}~\bibnamefont
  {Rubio}},\ }\bibfield  {title} {\enquote {\bibinfo {title} {Impact of the
  electronic band structure in high-harmonic generation spectra of solids},}\
  }\href {\doibase 10.1103/PhysRevLett.118.087403} {\bibfield  {journal}
  {\bibinfo  {journal} {Phys. Rev. Lett.}\ }\textbf {\bibinfo {volume} {118}},\
  \bibinfo {pages} {087403} (\bibinfo {year} {2017})}\BibitemShut {NoStop}%
\bibitem [{\citenamefont {Hansen}\ \emph {et~al.}(2017)\citenamefont {Hansen},
  \citenamefont {Deffge},\ and\ \citenamefont {Bauer}}]{Hansen17}%
  \BibitemOpen
  \bibfield  {author} {\bibinfo {author} {\bibfnamefont {K.~K.}\ \bibnamefont
  {Hansen}}, \bibinfo {author} {\bibfnamefont {T.}~\bibnamefont {Deffge}}, \
  and\ \bibinfo {author} {\bibfnamefont {D.}~\bibnamefont {Bauer}},\ }\bibfield
   {title} {\enquote {\bibinfo {title} {High-order harmonic generation in solid
  slabs beyond the single-active-electron approximation},}\ }\href {\doibase
  10.1103/PhysRevA.96.053418} {\bibfield  {journal} {\bibinfo  {journal} {Phys.
  Rev. A}\ }\textbf {\bibinfo {volume} {96}},\ \bibinfo {pages} {053418}
  (\bibinfo {year} {2017})}\BibitemShut {NoStop}%
\bibitem [{\citenamefont {Yu}\ \emph {et~al.}(2019)\citenamefont {Yu},
  \citenamefont {Jiang},\ and\ \citenamefont {Lu}}]{Yu19}%
  \BibitemOpen
  \bibfield  {author} {\bibinfo {author} {\bibfnamefont {Chao}\ \bibnamefont
  {Yu}}, \bibinfo {author} {\bibfnamefont {Shicheng}\ \bibnamefont {Jiang}}, \
  and\ \bibinfo {author} {\bibfnamefont {Ruifeng}\ \bibnamefont {Lu}},\
  }\bibfield  {title} {\enquote {\bibinfo {title} {High order harmonic
  generation in solids: a review on recent numerical methods},}\ }\href
  {\doibase 10.1080/23746149.2018.1562982} {\bibfield  {journal} {\bibinfo
  {journal} {Advances in Physics: X}\ }\textbf {\bibinfo {volume} {4}},\
  \bibinfo {pages} {1562982} (\bibinfo {year} {2019})}\BibitemShut {NoStop}%
\bibitem [{\citenamefont {Milo{\v{s}}evi{\'{c}}}\ and\ \citenamefont
  {Ehlotzky}(2003)}]{MEAdv03}%
  \BibitemOpen
  \bibfield  {author} {\bibinfo {author} {\bibfnamefont {D.~B.}\ \bibnamefont
  {Milo{\v{s}}evi{\'{c}}}}\ and\ \bibinfo {author} {\bibfnamefont
  {F.}~\bibnamefont {Ehlotzky}},\ }\bibfield  {title} {\enquote {\bibinfo
  {title} {Scattering and reaction processes in powerfull laser fields},}\
  }\href {\doibase https://doi.org/10.1016/S1049-250X(03)80007-1} {\bibfield
  {journal} {\bibinfo  {journal} {Adv. At., Mol., Opt. Phys.}\ }\textbf
  {\bibinfo {volume} {49}},\ \bibinfo {pages} {373} (\bibinfo {year}
  {2003})}\BibitemShut {NoStop}%
\bibitem [{\citenamefont {Sali{\` e}res}\ \emph {et~al.}(2001)\citenamefont
  {Sali{\` e}res}, \citenamefont {Carr{\' e}}, \citenamefont {Le~D{\' e}roff},
  \citenamefont {Grasbon}, \citenamefont {Paulus}, \citenamefont {Walther},
  \citenamefont {Kopold}, \citenamefont {Becker}, \citenamefont
  {Milo{\v{s}}evi{\'{c}}}, \citenamefont {Sanpera},\ and\ \citenamefont
  {Lewenstein}}]{SalScience01}%
  \BibitemOpen
  \bibfield  {author} {\bibinfo {author} {\bibfnamefont {P.}~\bibnamefont
  {Sali{\` e}res}}, \bibinfo {author} {\bibfnamefont {B.}~\bibnamefont {Carr{\'
  e}}}, \bibinfo {author} {\bibfnamefont {L.}~\bibnamefont {Le~D{\' e}roff}},
  \bibinfo {author} {\bibfnamefont {F.}~\bibnamefont {Grasbon}}, \bibinfo
  {author} {\bibfnamefont {G.~G.}\ \bibnamefont {Paulus}}, \bibinfo {author}
  {\bibfnamefont {H.}~\bibnamefont {Walther}}, \bibinfo {author} {\bibfnamefont
  {R.}~\bibnamefont {Kopold}}, \bibinfo {author} {\bibfnamefont
  {W.}~\bibnamefont {Becker}}, \bibinfo {author} {\bibfnamefont {D.~B.}\
  \bibnamefont {Milo{\v{s}}evi{\'{c}}}}, \bibinfo {author} {\bibfnamefont
  {A.}~\bibnamefont {Sanpera}}, \ and\ \bibinfo {author} {\bibfnamefont
  {M.}~\bibnamefont {Lewenstein}},\ }\bibfield  {title} {\enquote {\bibinfo
  {title} {Feynman's path-integral approach for intense-laser-atom
  interactions},}\ }\href {\doibase 10.1126/science.108836} {\bibfield
  {journal} {\bibinfo  {journal} {Science}\ }\textbf {\bibinfo {volume}
  {292}},\ \bibinfo {pages} {902} (\bibinfo {year} {2001})}\BibitemShut
  {NoStop}%
\bibitem [{\citenamefont {Levesque}\ \emph {et~al.}(2007)\citenamefont
  {Levesque}, \citenamefont {Zeidler}, \citenamefont {Marangos}, \citenamefont
  {Corkum},\ and\ \citenamefont {Villeneuve}}]{LZMCVPRL07}%
  \BibitemOpen
  \bibfield  {author} {\bibinfo {author} {\bibfnamefont {J.}~\bibnamefont
  {Levesque}}, \bibinfo {author} {\bibfnamefont {D.}~\bibnamefont {Zeidler}},
  \bibinfo {author} {\bibfnamefont {J.~P.}\ \bibnamefont {Marangos}}, \bibinfo
  {author} {\bibfnamefont {P.~B.}\ \bibnamefont {Corkum}}, \ and\ \bibinfo
  {author} {\bibfnamefont {D.~M.}\ \bibnamefont {Villeneuve}},\ }\bibfield
  {title} {\enquote {\bibinfo {title} {High harmonic generation and the role of
  atomic orbital wave functions},}\ }\href {\doibase
  10.1103/PhysRevLett.98.183903} {\bibfield  {journal} {\bibinfo  {journal}
  {Phys. Rev. Lett.}\ }\textbf {\bibinfo {volume} {98}},\ \bibinfo {pages}
  {183903} (\bibinfo {year} {2007})}\BibitemShut {NoStop}%
\bibitem [{\citenamefont {Okajima}\ \emph {et~al.}(2012)\citenamefont
  {Okajima}, \citenamefont {Tolstikhin},\ and\ \citenamefont
  {Morishita}}]{OTMPRA12}%
  \BibitemOpen
  \bibfield  {author} {\bibinfo {author} {\bibfnamefont {Y.}~\bibnamefont
  {Okajima}}, \bibinfo {author} {\bibfnamefont {O.~I.}\ \bibnamefont
  {Tolstikhin}}, \ and\ \bibinfo {author} {\bibfnamefont {T.}~\bibnamefont
  {Morishita}},\ }\bibfield  {title} {\enquote {\bibinfo {title} {Adiabatic
  theory of high-order harmonic generation: One-dimensional
  zero-range-potential model},}\ }\href
  {https://journals.aps.org/pra/abstract/10.1103/PhysRevA.85.063406} {\bibfield
   {journal} {\bibinfo  {journal} {Phys. Rev. A}\ }\textbf {\bibinfo {volume}
  {85}},\ \bibinfo {pages} {063406} (\bibinfo {year} {2012})}\BibitemShut
  {NoStop}%
\bibitem [{\citenamefont {Flegel}\ \emph {et~al.}(2021)\citenamefont {Flegel},
  \citenamefont {Manakov}, \citenamefont {Breev},\ and\ \citenamefont
  {Frolov}}]{FlegPRA21}%
  \BibitemOpen
  \bibfield  {author} {\bibinfo {author} {\bibfnamefont {A.~V.}\ \bibnamefont
  {Flegel}}, \bibinfo {author} {\bibfnamefont {N.~L.}\ \bibnamefont {Manakov}},
  \bibinfo {author} {\bibfnamefont {Ia.~V.}\ \bibnamefont {Breev}}, \ and\
  \bibinfo {author} {\bibfnamefont {M.~V.}\ \bibnamefont {Frolov}},\ }\bibfield
   {title} {\enquote {\bibinfo {title} {Adiabatic expressions for the wave
  function of an electron in a finite-range potential and an intense
  low-frequency laser pulse},}\ }\href {\doibase 10.1103/PhysRevA.104.033109}
  {\bibfield  {journal} {\bibinfo  {journal} {Phys. Rev. A}\ }\textbf {\bibinfo
  {volume} {104}},\ \bibinfo {pages} {033109} (\bibinfo {year}
  {2021})}\BibitemShut {NoStop}%
\bibitem [{\citenamefont {Flegel}\ and\ \citenamefont
  {Frolov}(2023)}]{FrolovJPA2023}%
  \BibitemOpen
  \bibfield  {author} {\bibinfo {author} {\bibfnamefont {A.~V.}\ \bibnamefont
  {Flegel}}\ and\ \bibinfo {author} {\bibfnamefont {M.~V.}\ \bibnamefont
  {Frolov}},\ }\bibfield  {title} {\enquote {\bibinfo {title} {On the adiabatic
  approximation for the laser-dressed atom},}\ }\href {\doibase
  10.1088/1751-8121/ad0dc8} {\bibfield  {journal} {\bibinfo  {journal} {J.
  Phys. A: Math. Theor.}\ }\textbf {\bibinfo {volume} {50}},\ \bibinfo {pages}
  {505304} (\bibinfo {year} {2023})}\BibitemShut {NoStop}%
\bibitem [{\citenamefont {Frolov}\ \emph {et~al.}(2009)\citenamefont {Frolov},
  \citenamefont {Manakov}, \citenamefont {Sarantseva}, \citenamefont {Emelin},
  \citenamefont {Ryabikin},\ and\ \citenamefont {Starace}}]{FMSERSPRL09}%
  \BibitemOpen
  \bibfield  {author} {\bibinfo {author} {\bibfnamefont {M.~V.}\ \bibnamefont
  {Frolov}}, \bibinfo {author} {\bibfnamefont {N.~L.}\ \bibnamefont {Manakov}},
  \bibinfo {author} {\bibfnamefont {T.~S.}\ \bibnamefont {Sarantseva}},
  \bibinfo {author} {\bibfnamefont {M.~{Yu}.}\ \bibnamefont {Emelin}}, \bibinfo
  {author} {\bibfnamefont {M.~{Yu}.}\ \bibnamefont {Ryabikin}}, \ and\ \bibinfo
  {author} {\bibfnamefont {A.~F.}\ \bibnamefont {Starace}},\ }\bibfield
  {title} {\enquote {\bibinfo {title} {Analytic description of the high-energy
  plateau in harmonic generation by atoms: {C}an the harmonic power increase
  with increasing laser wavelengths?}}\ }\href {\doibase
  10.1103/PhysRevLett.102.243901} {\bibfield  {journal} {\bibinfo  {journal}
  {Phys. Rev. Lett.}\ }\textbf {\bibinfo {volume} {102}},\ \bibinfo {pages}
  {243901} (\bibinfo {year} {2009})}\BibitemShut {NoStop}%
\bibitem [{\citenamefont {Frolov}\ \emph {et~al.}(2010)\citenamefont {Frolov},
  \citenamefont {Manakov},\ and\ \citenamefont {Starace}}]{FMSPRA10}%
  \BibitemOpen
  \bibfield  {author} {\bibinfo {author} {\bibfnamefont {M.~V.}\ \bibnamefont
  {Frolov}}, \bibinfo {author} {\bibfnamefont {N.~L.}\ \bibnamefont {Manakov}},
  \ and\ \bibinfo {author} {\bibfnamefont {A.~F.}\ \bibnamefont {Starace}},\
  }\bibfield  {title} {\enquote {\bibinfo {title} {Potential barrier effects in
  high-order harmonic generation by transition-metal ions},}\ }\href@noop {}
  {\bibfield  {journal} {\bibinfo  {journal} {Phys. Rev. A}\ }\textbf {\bibinfo
  {volume} {82}},\ \bibinfo {pages} {023424} (\bibinfo {year}
  {2010})}\BibitemShut {NoStop}%
\bibitem [{\citenamefont {Shiner}\ \emph {et~al.}(2011)\citenamefont {Shiner},
  \citenamefont {Schmidt}, \citenamefont {Trallero-Herrero}, \citenamefont
  {W\"{o}rner}, \citenamefont {Patchkovskii}, \citenamefont {Corkum},
  \citenamefont {Kieffer}, \citenamefont {L\'{e}gar\'{e}},\ and\ \citenamefont
  {Villeneuve}}]{NatureTralerro}%
  \BibitemOpen
  \bibfield  {author} {\bibinfo {author} {\bibfnamefont {A.~D.}\ \bibnamefont
  {Shiner}}, \bibinfo {author} {\bibfnamefont {B.~E.}\ \bibnamefont {Schmidt}},
  \bibinfo {author} {\bibfnamefont {C.}~\bibnamefont {Trallero-Herrero}},
  \bibinfo {author} {\bibfnamefont {H.~J.}\ \bibnamefont {W\"{o}rner}},
  \bibinfo {author} {\bibfnamefont {S.}~\bibnamefont {Patchkovskii}}, \bibinfo
  {author} {\bibfnamefont {P.~B.}\ \bibnamefont {Corkum}}, \bibinfo {author}
  {\bibfnamefont {{\relax J.--C}.}~\bibnamefont {Kieffer}}, \bibinfo {author}
  {\bibfnamefont {F.}~\bibnamefont {L\'{e}gar\'{e}}}, \ and\ \bibinfo {author}
  {\bibfnamefont {D.~M.}\ \bibnamefont {Villeneuve}},\ }\bibfield  {title}
  {\enquote {\bibinfo {title} {Probing collective multi-electron dynamics in
  xenon with high-harmonic spectroscopy},}\ }\href {\doibase
  https://doi.org/10.1038/nphys1940} {\bibfield  {journal} {\bibinfo  {journal}
  {Nat. Phys.}\ }\textbf {\bibinfo {volume} {7}},\ \bibinfo {pages} {464}
  (\bibinfo {year} {2011})}\BibitemShut {NoStop}%
\bibitem [{\citenamefont {Shiner}\ \emph {et~al.}(2012)\citenamefont {Shiner},
  \citenamefont {Schmidt}, \citenamefont {Trallero-Herrero}, \citenamefont
  {Corkum}, \citenamefont {Kieffer}, \citenamefont {L\'{e}gar\'{e}},\ and\
  \citenamefont {Villeneuve}}]{SSetalJPB12}%
  \BibitemOpen
  \bibfield  {author} {\bibinfo {author} {\bibfnamefont {A.~D.}\ \bibnamefont
  {Shiner}}, \bibinfo {author} {\bibfnamefont {B.~E.}\ \bibnamefont {Schmidt}},
  \bibinfo {author} {\bibfnamefont {C.}~\bibnamefont {Trallero-Herrero}},
  \bibinfo {author} {\bibfnamefont {P.~B.}\ \bibnamefont {Corkum}}, \bibinfo
  {author} {\bibfnamefont {J.-C.}\ \bibnamefont {Kieffer}}, \bibinfo {author}
  {\bibfnamefont {F.}~\bibnamefont {L\'{e}gar\'{e}}}, \ and\ \bibinfo {author}
  {\bibfnamefont {D.~M.}\ \bibnamefont {Villeneuve}},\ }\bibfield  {title}
  {\enquote {\bibinfo {title} {Observation of cooper minimum in krypton using
  high harmonic spectroscopy},}\ }\href
  {http://iopscience.iop.org/article/10.1088/0953-4075/45/7/074010/meta}
  {\bibfield  {journal} {\bibinfo  {journal} {J. Phys. B: At. Mol. Opt. Phys.}\
  }\textbf {\bibinfo {volume} {45}},\ \bibinfo {pages} {074010} (\bibinfo
  {year} {2012})}\BibitemShut {NoStop}%
\bibitem [{\citenamefont {Romanov}\ \emph {et~al.}(2020)\citenamefont
  {Romanov}, \citenamefont {Silaev}, \citenamefont {Frolov},\ and\
  \citenamefont {Vvedenskii}}]{SilaevPRA20}%
  \BibitemOpen
  \bibfield  {author} {\bibinfo {author} {\bibfnamefont {A.~A.}\ \bibnamefont
  {Romanov}}, \bibinfo {author} {\bibfnamefont {A.~A.}\ \bibnamefont {Silaev}},
  \bibinfo {author} {\bibfnamefont {M.~V.}\ \bibnamefont {Frolov}}, \ and\
  \bibinfo {author} {\bibfnamefont {N.~V.}\ \bibnamefont {Vvedenskii}},\
  }\bibfield  {title} {\enquote {\bibinfo {title} {Influence of the
  polarization of a multielectron atom in a strong laser field on high-order
  harmonic generation},}\ }\href {\doibase 10.1103/PhysRevA.101.013435}
  {\bibfield  {journal} {\bibinfo  {journal} {Phys. Rev. A}\ }\textbf {\bibinfo
  {volume} {101}},\ \bibinfo {pages} {013435} (\bibinfo {year}
  {2020})}\BibitemShut {NoStop}%
\bibitem [{\citenamefont {Romanov}\ \emph {et~al.}(2021)\citenamefont
  {Romanov}, \citenamefont {Silaev}, \citenamefont {Sarantseva}, \citenamefont
  {Frolov},\ and\ \citenamefont {Vvedenskii}}]{Romanov2021}%
  \BibitemOpen
  \bibfield  {author} {\bibinfo {author} {\bibfnamefont {A.~A.}\ \bibnamefont
  {Romanov}}, \bibinfo {author} {\bibfnamefont {A.~A.}\ \bibnamefont {Silaev}},
  \bibinfo {author} {\bibfnamefont {T.~S.}\ \bibnamefont {Sarantseva}},
  \bibinfo {author} {\bibfnamefont {M.~V.}\ \bibnamefont {Frolov}}, \ and\
  \bibinfo {author} {\bibfnamefont {N.~V.}\ \bibnamefont {Vvedenskii}},\
  }\bibfield  {title} {\enquote {\bibinfo {title} {Study of high-order harmonic
  generation in xenon based on time-dependent density-functional theory},}\
  }\href {\doibase 10.1088/1367-2630/abe8a9} {\bibfield  {journal} {\bibinfo
  {journal} {New J. Phys.}\ }\textbf {\bibinfo {volume} {23}},\ \bibinfo
  {pages} {043014} (\bibinfo {year} {2021})}\BibitemShut {NoStop}%
\bibitem [{\citenamefont {Krieger}\ and\ \citenamefont
  {Iafrate}(1986)}]{Krieger86}%
  \BibitemOpen
  \bibfield  {author} {\bibinfo {author} {\bibfnamefont {J.~B.}\ \bibnamefont
  {Krieger}}\ and\ \bibinfo {author} {\bibfnamefont {G.~J.}\ \bibnamefont
  {Iafrate}},\ }\bibfield  {title} {\enquote {\bibinfo {title} {Time evolution
  of bloch electrons in a homogeneous electric field},}\ }\href {\doibase
  10.1103/PhysRevB.33.5494} {\bibfield  {journal} {\bibinfo  {journal} {Phys.
  Rev. B}\ }\textbf {\bibinfo {volume} {33}},\ \bibinfo {pages} {5494--5500}
  (\bibinfo {year} {1986})}\BibitemShut {NoStop}%
\bibitem [{\citenamefont {Zener}(1934)}]{Zener34}%
  \BibitemOpen
  \bibfield  {author} {\bibinfo {author} {\bibfnamefont {C}~\bibnamefont
  {Zener}},\ }\bibfield  {title} {\enquote {\bibinfo {title} {A theory of the
  electrical breakdown of solid dielectris},}\ }\href {\doibase
  10.1098/rspa.1934.0116} {\bibfield  {journal} {\bibinfo  {journal} {Proc. R.
  Soc. Lond. A}\ }\textbf {\bibinfo {volume} {145}},\ \bibinfo {pages}
  {523--529} (\bibinfo {year} {1934})}\BibitemShut {NoStop}%
\bibitem [{\citenamefont {Landau}\ and\ \citenamefont {Lifshitz}(1977)}]{LL3}%
  \BibitemOpen
  \bibfield  {author} {\bibinfo {author} {\bibfnamefont {L.~D.}\ \bibnamefont
  {Landau}}\ and\ \bibinfo {author} {\bibfnamefont {E.~M.}\ \bibnamefont
  {Lifshitz}},\ }\href@noop {} {\emph {\bibinfo {title} {Quantum Mechanics
  (Non-relativistic Theory)}}}\ (\bibinfo  {publisher} {Pergamon Press,
  Oxford},\ \bibinfo {year} {1977})\BibitemShut {NoStop}%
\end{thebibliography}
%

\end{document}